\documentclass[10pt,journal,compsoc]{IEEEtran}

\usepackage{amsmath,amsfonts}
\usepackage{textcomp}
\usepackage{stfloats}
\usepackage{url}
\usepackage{verbatim}
\usepackage{graphicx}

\ifCLASSOPTIONcompsoc
  \usepackage[nocompress]{cite}
\else
  \usepackage{cite}
\fi

\usepackage[utf8]{inputenc} %
\usepackage{hyperref}       %
\usepackage{url}            %
\usepackage{booktabs}       %
\usepackage{amsfonts}       %
\usepackage{nicefrac}       %
\usepackage{microtype}      %
\usepackage{xcolor}         %

\usepackage{booktabs} %

\usepackage{mathrsfs}
\usepackage{caption}
\usepackage{setspace}
\usepackage[noend]{algpseudocode}
\usepackage{graphicx}
\usepackage{enumerate}
\usepackage{bm}
\usepackage{multirow}
\usepackage[flushleft]{threeparttable}
\usepackage{enumitem}
\usepackage{soul}
\usepackage[normalem]{ulem}
\usepackage[ruled,linesnumbered]{algorithm2e}
\usepackage{subfigure}
\usepackage{amsthm}
\usepackage{ifthen}

\newcommand{\tabincell}[2]{\begin{tabular}{@{}#1@{}}#2\end{tabular}}

\def\BibTeX{{\rm B\kern-.05em{\sc i\kern-.025em b}\kern-.08em
    T\kern-.1667em\lower.7ex\hbox{E}\kern-.125emX}}
\usepackage{balance}

\begin{document}
\title{i-Razor: A Differentiable Neural Input Razor for Feature Selection and Dimension Search in DNN-Based Recommender Systems}
\author{Yao~Yao, %
        Bin~Liu, %
        Haoxun He, 
        Dakui Sheng, 
        Ke Wang, 
        Li Xiao, 
        and Huanhuan Cao
\IEEEcompsocitemizethanks{\IEEEcompsocthanksitem Yao Yao and Li Xiao are with the Tsinghua-Berkeley Shenzhen Institute (TBSI), Tsinghua University, Shenzhen, Guangdong, 518055, China.\protect\\
E-mail:  y-yao19@mails.tsinghua.edu.cn, xiaoli@sz.tsinghua.edu.cn.
\IEEEcompsocthanksitem Bin Liu, Haoxun He, Dakui Sheng, Ke Wang, and Huanhuan Cao are with ByteDance Inc, Beijing, 100098, China.\protect\\
E-mail:  liubinbh@126.com, \{hehaoxun, shengdakui, wangke, caohuanhuan\}@bytedance.com.}
\thanks{Manuscript submitted August 3, 2022.\\(Corresponding author: Li Xiao.)}}

\markboth{Journal of \LaTeX\ Class Files,~Vol.~18, No.~9, September~2020}%
{How to Use the IEEEtran \LaTeX \ Templates}

\IEEEtitleabstractindextext{%
\begin{abstract}
Input features play a crucial role in DNN-based recommender systems with thousands of categorical and continuous fields from users, items, contexts, and interactions. Noisy features and inappropriate embedding dimension assignments can deteriorate the performance of recommender systems and introduce unnecessary complexity in model training and online serving. Optimizing the input configuration of DNN models, including feature selection and embedding dimension assignment, has become one of the essential topics in feature engineering. However, in existing industrial practices, feature selection and dimension search are optimized sequentially, i.e., feature selection is performed first, followed by dimension search to determine the optimal dimension size for each selected feature. Such a sequential optimization mechanism increases training costs and risks generating suboptimal input configurations. To address this problem, we propose a differentiable neural \textbf{i}nput \textbf{razor} (\textbf{i-Razor}) that enables joint optimization of feature selection and dimension search. Concretely, we introduce an end-to-end differentiable model to learn the relative importance of different embedding regions of each feature. Furthermore, a flexible pruning algorithm is proposed to achieve feature filtering and dimension derivation simultaneously. Extensive experiments on two large-scale public datasets in the Click-Through-Rate (CTR) prediction task demonstrate the efficacy and superiority of i-Razor in balancing model complexity and performance.
\end{abstract}

}
\maketitle
\IEEEdisplaynontitleabstractindextext
\IEEEpeerreviewmaketitle

\IEEEraisesectionheading{\section{Introduction}\label{sec:intro}}

\begin{figure}[t!]
    \centering
    \includegraphics[width=0.48\textwidth]{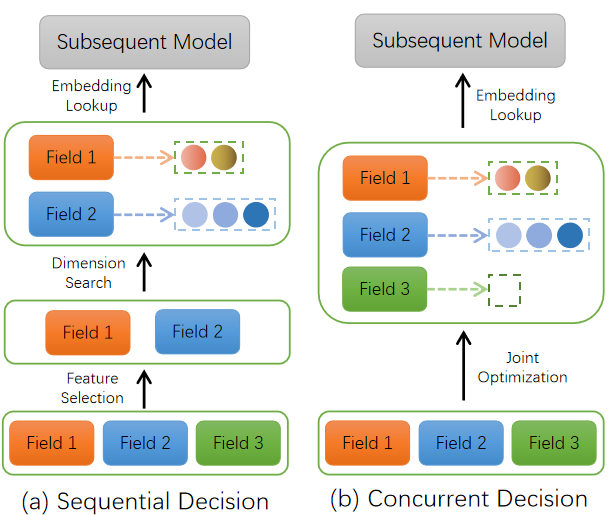}
    \caption{{Contrasting two strategies for optimizing model inputs. (a) Sequentially conducting feature selection and tuning feature dimensions. (b) Simultaneously optimizing feature selection and dimension allocation.}}
    \label{fig:optima}
\end{figure}

\IEEEPARstart{I}{n} this era of information overload, recommender systems powered by information technology and deep learning have become an effective way to retrieve potentially useful information for users from a huge range of options~\cite{pazzani2007content,chen2022clcdr,chen2023cdr}.
Deep learning recommender systems enhance recommendation performance by capturing complex correlations between features~\cite{autocross}.
Experts resort to feature engineering, generating categorical, numerical, statistical, and cross features, to better understand the user's interests.
However, due to the universal approximation property~\cite{hornik1989multilayer} of deep neural networks (DNN), feeding noisy features into a DNN model can adversely affect predictive performance~\cite{autocross,muthusankar2019high}.
Meanwhile, because the embedding lookup table dominates both the size and inductive bias of DNN-based recommendation models~\cite{cheng2020differentiable}, the embedding dimensions of input features are crucial to the performance and complexity of recommender systems.
Therefore, it is highly desired to efficiently identify effective features from the original feature set and specify appropriate embedding dimensions.
{As illustrated in Figure~\ref{fig:optima}, conventional industrial practices typically optimize feature selection and embedding dimension allocation separately. In contrast, this work explores jointly optimizing both in an end-to-end fashion. }

{Feature selection is a crucial task in recommender systems and has been the subject of extensive research. Previous studies have categorized feature selection methods into three main categories: \textit{Filter}, \textit{Wrapper}, and \textit{Embedded} methods~\cite{chandrashekar2014survey,wang2020bacterial,zheng2023automl}.}
\textit{Filter} methods measure the discriminant attributes of features through criteria such as information gain~\cite{wang2011igf} and feature consistency~\cite{dash2003consistency}, so as to extract high-ranked features. 
{\textit{Wrapper} methods optimize feature subsets by using performance assessments such as the K-nearest neighbor (KNN) algorithm and linear discriminant analysis (LDA) instead of solely relying on evaluation criteria~\cite{xue2015survey}.}
Compared with \textit{Filter} and \textit{Wrapper}, \textit{Embedded} methods, such as the least absolute shrinkage and selection operator (LASSO)~\cite{liu2015pairwise} and the gradient-boosting decision tree (GBDT)~\cite{friedman2001greedy,he2014practical}, incorporate feature selection as part of the training process. Despite compensating for the drawbacks of low efficiency in \textit{Filter} and high computational cost in \textit{Wrapper}, the effectiveness of \textit{Embedded} methods is substantially dependent on conductive parameters and strict model assumptions~\cite{wang2020bacterial}. 
Although the aforementioned feature selection methods have proven effective in several scenarios,
{there could be a mismatch between the subset of features selected to optimize the feature selection model and the subset that would ideally serve the target DNN model.} 
{Recognizing that DNN models can capture more complex feature interactions, it would be logical to use the target DNN model to guide feature selection directly.} 
{In light of this, we propose a model-consistent feature selection approach that integrates the target DNN model into the feature selection process.} 
Our approach belongs to the \textit{Embedded} methods in a broad sense, but there is a notable difference between our study and previous work: we evaluate the importance of features from the point of view of each feature's optimal embedding dimension size under the target DNN model.

In recent years, there has been an increasing interest in studying embedding dimensions, as the embedding layer impacts on both the scale and the inductive bias of DNN-based recommendation models~\cite{cheng2020differentiable}. Specifying an adaptive embedding dimension for each feature is a problem of exponential complexity. A common practice is to set a unified value for all features by hyperparameter tuning experiments or relying on expert knowledge. However, such practice does not distinguish between different features and can be memory inefficient.
Inspired by recent progress in neural architecture search (NAS)~\cite{liu2018darts}, methods based on reinforcement learning~\cite{joglekar2020neural}, differentiable search~\cite{zhao2020memory,zhao2020autoemb}, pruning~\cite{liu2021learnable}, and so forth, have been proposed to perform dimension search at either field level or feature level~\footnote{Field-level methods assign the same dimension to all features within the same field while feature-level ones can assign diverse dimensions to different features within the same field.}.
{Despite the ability of these methods to reduce model parameters without compromising model performance given a carefully selected feature subset, their effectiveness without feature selection remains unexplored. The dependency on feature selection effectively equates to assuming all input features are beneficial during the dimension search process.} 
{We believe that there is a necessity to conduct dimension search under the assumption that not all features in the original feature set are helpful, as the updating of features can be very frequent in industrial practice. With this objective, we propose a differentiable framework that identifies redundantly assigned embedding dimensions for pruning.}
{Specifically, a feature with its embedding dimension pruned to 0 is considered unimportant. It is worth noting that features are typically grouped into fields for practical reasons in industrial recommender systems, and these fields can have significantly varying cardinalities.}
For example, the field \textit{Gender} generally contains two features ``Female'' and ``Male'',  while the field \textit{Browsed Goods} may have millions of features.
{While individual dimension assignment for each feature (i.e., feature-level methods) might boost performance, field selection and field-level dimension search are often prioritized in large-scale recommender systems to reduce complexity and enhance implementation ease.} 
{This paper emphasizes assigning different dimension sizes to fields based on their importance, and discarding less important fields during the retraining stage. Unless otherwise specified, the term \textbf{``feature selection''} hereafter refers to the selection of fields.} 

To guide the input configuration towards better optima, we propose a differentiable neural \textbf{i}nput \textbf{razor} (\textbf{i-Razor}) that is capable of simultaneously pruning unnecessary fields and assigning adaptive embedding dimensions. 
The training of i-Razor consists of two stages. \romannumeral1) In the pretraining stage, we refine the dimension search space to learn the importance of different fields and different embedding regions. Meanwhile, to control the number of model parameters and avoid overfitting, we introduce a novel regularization term to regularize the model complexity. \romannumeral2) In the retraining stage, a proposed pruning algorithm is executed to remove redundant fields and preserve the desired embedding dimensions for the remaining fields according to the relative weights of each embedding region. We then reconstruct the DNN model and retrain it with the training dataset based on the derived input configuration.
We conduct extensive experiments on two large-scale public datasets for the CTR prediction task.
The experimental results show that i-Razor achieves higher prediction accuracy while saving fields than several state-of-the-art methods.

The main contributions of this paper are threefold.
\begin{itemize}
\item {
Traditionally, feature selection and embedding dimension search were treated as separate processes. Our research presents a pioneering approach that explores their joint optimization. This innovative leap  not only bridges the gap between feature selection and dimension search but also hints at a higher level of alignment between the two, which has the potential to discover input configurations that more closely approach optimal solutions.} 
\item {We put forth i-Razor, a new paradigm for achieving model-consistent feature selection, with dimension search as the conduit. Distinct from conventional dimension search methods, i-Razor dispenses with the prerequisite for a pre-selected set of features. By integrating dimension space refinement and a novel pruning algorithm, it concurrently conducts feature selection and dimension search, thereby diminishing the total training cost of the input configuration optimization process.}
\item We have validated the effectiveness of i-Razor through comprehensive experiments on two large-scale public datasets. The empirical results suggest that i-Razor is capable of removing unnecessary fields  and compressing redundant embedding dimensions, without compromising model performance. 
{Consequently, it contributes to significant savings in parameter and storage overheads of the online serving model, establishing i-Razor's practical value.} 
\end{itemize}

The remainder of this paper is organized as follows: Section~\ref{sec:related} discusses the related literature on feature selection and embedding dimension search. We formulate the input configuration optimization problem and present the framework of i-Razor in Section~\ref{sec:method}. In Section~\ref{sec:offline-exp}, we elaborate on our experimental setup and empirical results. Finally, we conclude our work and discuss its potential future extensions in Section~\ref{sec:conclusion}.

\section{Related Work}\label{sec:related}
In this section, we briefly review the works related to our study, including two branches of the literature: feature selection approaches and dimension search techniques.
\subsection{Feature Selection Approaches}
Feature selection refers to the process of finding out a subset from the original feature set without losing useful information~\cite{cai2018feature}. 
Existing feature selection methods include the following three categories: \textit{Filter}, \textit{Wrapper}, and \textit{Embedded} methods~\cite{mafarja2020dragonfly}. \textit{Filter} methods measure the association between the feature and the class label via some specific evaluation criteria~\cite{cai2018feature}, such as information gain~\cite{lee2006information}, minimum redundant maximum relevance~\cite{peng2005feature},  {and spatio-temporal preserving representations~\cite{zhang2019making}}. This stream of research also includes feature selection methods based on statistical tests, e.g., Chi-squared~\cite{1996A}, Pearson correlation, and ANOVA F-value test~\cite{1965An}. 
Input features that {demonstrate a significant} 
statistical relationship with the output variable  {are}  retained and utilized for further analysis,  {while others are discarded.} 
\textit{Filter} methods are expected to have the highest computational efficiency and tend to keep more features from the total feature space{~\cite{wang2020bacterial,zheng2023automl}.}
 {By contrast, \textit{Wrapper} methods adopt specific learning techniques such as} 
 KNN~\cite{dudani1976distance}, Naive Bayes~\cite{rish2001empirical}, LDA~\cite{ahdesmaki2010feature},  {and adaptive semi-supervised feature analysis~\cite{luo2017adaptive}} to assess the performance of feature combinations~\cite{xue2015survey}. 
 {Though more efficient for feature selection than \textit{Filter} methods, \textit{Wrapper} methods require higher computational resources}~\cite{pourpanah2019hybrid}.
In these two types of feature selection methods, the feature selection process and the model training process are distinctly separate. 

Unlike \textit{Filter} and \textit{Wrapper} methods, \textit{Embedded} methods such as LASSO~\cite{liu2015pairwise}, GBDT~\cite{friedman2001greedy},  {and semi-supervised recurrent convolutional attention models~\cite{chen2019semisupervised}} integrate the feature selection process with the model training process, i.e., feature selection is performed automatically during the model training process~\cite{chandrashekar2014survey}. 
 {Recently, the advancement of NAS has spawned a new research hotspot in \textit{Embedded} methods, the automatic search for useful features with AutoML techniques~\cite{zheng2023automl,cheng2022towards}.}
For instance, AutoFIS~\cite{liu2020autofis} proposes a gating mechanism to identify and mask the intersection of useless features in factorization models.
 {AutoField~\cite{wang2022autofield} assigns two operators to each field, one for selecting and the other for ignoring that field. By comparing the relative importance of these two operators, it determines whether to select that field or not. 
Operators of each field of AutoField~\cite{wang2022autofield} and AutoFIS~\cite{liu2020autofis} are independent of each other, while AdaFS~\cite{lin2022adafs} assigns associated operators to each field through softmax operation, and measures the relative importance of fields corresponding to each operator by comparing the weight of each operator, so as to screen features.
}
 {Meanwhile, there has been a growing interest in automatic feature selection in deep learning-based sequential recommendation systems. For instance, NASR~\cite{cheng2022towards} made significant strides in designing a hybrid sequential recommendation model with the aim to ensemble the capacity of both self-attention and convolutional architectures. The proposed method can automatically select the architecture operation on each layer.}

 {Inspired by these advancements, our approach, i-Razor, also belongs to the \textit{Embedded} category, but with a unique perspective. i-Razor facilitates model-consistent feature selection by evaluating the importance of each feature concerning its optimal embedding dimension in the target DNN model. Building upon the foundation laid by AutoField, we propose a 0-dimension absorbing operator, which signifies the action of deselecting a specific feature. In addition, we enhance the existing design of the dimension search space, which commonly segments the embedding dimensions, by introducing the concept of the 0-dimension operator into this process. The weight associated with each operator measures the relative importance of these divisions, with 0 dimension indicating that all regions should be masked. By considering combinations of these operators, we derive the final input configuration. This method allows us to perform dimension search and feature selection concurrently, offering a more efficient manner to ensure that the chosen features and embedding dimensions contribute to the target model's performance.}

\subsection{Dimension Search Techniques}
Since assigning unified embedding dimensions to all features is incompatible with their heterogeneity, the new paradigm of mixed dimensions has gained more and more attention~\cite{ginart2019mixed,zhao2020autoemb,cheng2020differentiable,joglekar2020neural}. 
To be specific, in a mixed dimension scheme, different features can have different embedding dimensions. 
 There are two types of dimension search methods:
feature-level and field-level. 
Feature-level methods aim to assign different embedding dimensions to different features within the same field. For example, NIS~\cite{joglekar2020neural} applies reinforcement learning to generate decision sequences on the selection of dimensions for different features. Other heuristic approaches, such as MDE~\cite{ginart2019mixed}, DPG~\cite{chen2020differentiable}, and MGQE~\cite{kang2020learning}, attempt to assign higher dimensions to features with higher frequencies. However, these methods suffer from numerous unique values in each field, while the feature frequencies are highly dynamic and not pre-known in real-life situations. Although the frequency of different features can be estimated by sampling, the fine-grained dimension search at feature level introduces a sizeable computational overhead and additional complexity. In this work, we focus on the second group for its efficiency, which aims to allocate different dimensions to different fields, while different features in the same field share a common dimension size. 
Based on advances in NAS~\cite{mei2019atomnas}, the second type of dimension search method normally adopts an AutoML style to search for the desired embedding dimension from a set of predefined candidate embedding dimensions. 
Inspired by DARTS~\cite{liang2019darts+,liu2018darts}, AutoDim~\cite{zhao2020memory} proposes a differentiable search method to analyze the suitability of different candidate embedding dimensions by calculating the attention coefficients of the corresponding operators. 
After the search, AutoDim assigns to each field the candidate dimension size whose corresponding operator has maximum attention.
DNIS~\cite{cheng2020differentiable} eliminates the dependency on predefined candidate dimensions by directly pruning non-informative embedding blocks whose values are less than a given threshold. 
Although existing dimension search methods work well in the case of a well-selected subset of fields, their effectiveness in the presence of the original set of fields without feature selection remains understudied.

Similar to AutoDim, our objective is to automatically search for the optimal embedding dimension configuration on a given set of embedding dimension candidates. We differ from AutoDim: \romannumeral1) We consider dimension search on the original field set and avoid dimension assignment to useless fields by introducing a novel candidate dimension, i.e., dimension \textbf{0}. 
After the search is completed, the redundant dimensions and the fields dominated by dimension \textbf{0} will be discarded, which means that dimension search and feature selection can be achieved simultaneously. 
\romannumeral2) There is overlap between different candidate dimensions in AutoDim, i.e., the larger embedding dimensions cover the smaller ones, whereas the candidate dimensions in i-Razor do not overlap with each other. \romannumeral3) Unlike AutoDim, which uses the argmax operator for dimension pruning, we propose a flexible pruning algorithm that allows more fine-grained dimension assignment.

\begin{figure*}[t!]
  \centering
  \includegraphics[width=1\textwidth]{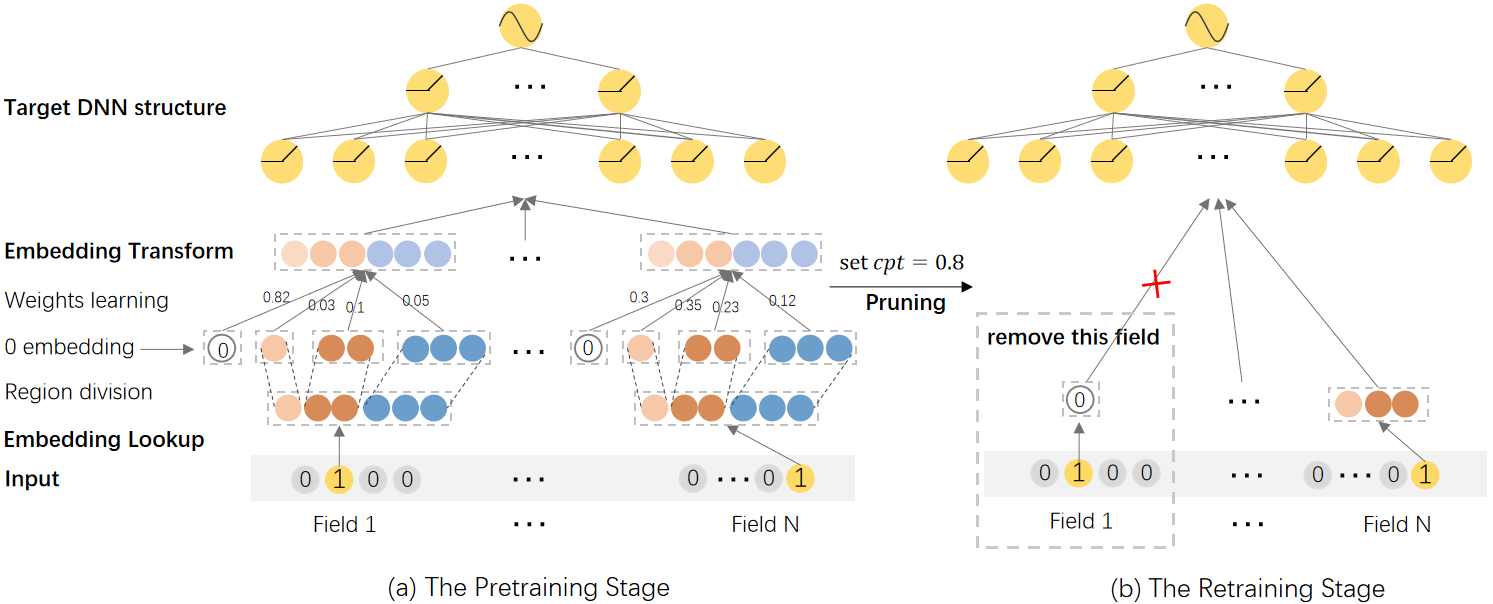}
  \caption{The overall framework of i-Razor for joint optimization of feature selection and dimension search.
  (a) In the pretraining stage, we divide the embedding of each field into several parts and learn the relative importance of these regions by training directly under the guidance of the target model. (b) In the retraining stage, we use a pruning algorithm to filter the useless fields and keep only the desired dimensions to reconstruct the model. After retraining it with the training dataset, the ultimate serving model can be obtained.}
  \label{fig:framework}
\end{figure*}

\section{Methodology}
\label{sec:method}
In this section, we first formulate the feature selection and embedding dimension search problem. %
Then, we describe the proposed i-Razor, a two-stage end-to-end framework to automatically select important fields and assign adaptive dimensions to the selected fields.

\subsection{Problem Formulation}
We assume that the original field set $\mathcal{F}$ consists of $N$ fields, whether useful or not.
Meanwhile, taking online resources into consideration, we can roughly give the search space of dimension search: $\mathcal{D}=\{d_1, \cdots, d_K\}$, where $K$ represents the number of candidate dimensions and $d_1< \cdots<d_K$. 
A solution $\mathcal{A}$ to the input configuration optimization problem involves not only filtering a field subset $\widetilde{\mathcal{F}} \subseteq \mathcal{F}$ in favor of the target DNN model, but also specifying the corresponding embedding dimension size for these selected fields.
Denoting the configuration of $\mathcal{A}$ by $\mathcal{C}_\mathcal{A} = \{(f_i, {d}_{f_i}) \mid f_i \in  \widetilde{\mathcal{F}}\}$, we can formulate the problem of finding the configuration that maximizes the performance of the target DNN model as follows:

\begin{equation}
\label{def_pro}
    \mathop{min}\limits_{\mathcal{A}} \mathcal{L}\left(\mathcal{C}_\mathcal{A}, \mathcal{W}\left(\mathcal{A}\right)\right),
\end{equation} 
where $\mathcal{L}$ is a task-specific loss function, $\mathcal{W}(\mathcal{A})$ denotes the optimal parameters of the DNN model under configuration $\mathcal{A}$, and ${d}_{f_i}$ specifies the unified dimension size assigned to all features in field $f_i$.

Finding the optimal solution to the input configuration optimization problem is an NP-Hard problem, with an incredibly huge space (typical size of $2^{NK}$) to search for.
For the purpose of automatically selecting essential features from a noisy feature set while learning proper dimensions simultaneously without loss of generality and model performance, we propose an end-to-end differentiable framework, i-Razor. The overall framework is illustrated in Figure \ref{fig:framework}, which is made up of two stages:
the {pretraining stage} and the {retraining stage}. 
We design the search space and the differentiable architecture for dimension search in the {pretraining stage}. To be specific, after an embedding layer, we adopt the target DNN structure to get model-consistent field importance evaluation by learning the relative weights of different operators. After that, in the {retraining stage}, a pruning algorithm is proposed to abandon unnecessary fields and determine the dimensions of the selected fields so that the target model can be retrained with the ultimate configuration.

\subsection{Pretraining Stage}
The input search problem is actually consistent with hyperparameter optimization \cite{franceschi2018bilevel} in a broad sense, since the decision of the field subset $\widetilde{\mathcal{F}}$ and the corresponding dimension configuration $\mathcal{C}_\mathcal{A}$ can be regarded as model hyperparameters to be determined before retraining the target model. 
However, the main difference is that the search space of $\mathcal{C}_\mathcal{A}$ in our problem is much larger than that of conventional hyperparameter optimization problems.

\subsubsection{Search Space}
\label{sec:Search space}
Before performing the input search, a suitable search space should be given in advance for dimension search.
Generally, previous work \cite{zhao2020memory,cheng2020differentiable} defined the dimension candidate set $\mathcal{D}$ as a set of positive integers, which implicitly assumes that every field is indispensable to the performance of the model. The assumption holds well under fine-grained feature engineering. However, in industrial recommender systems, features are iterated rapidly to continuously improve the user experience, 
{but not all newly generated features/fields are beneficial.} 
{When only given the original field set $\mathcal{F}$ rather than a well-chosen subset $\widetilde{\mathcal{F}}\subseteq \mathcal{F}$, allocating dimensions to noisy fields would introduce superfluous memory consumption and lead to suboptimal performance. To avoid forcing dimension assignment to noisy fields, we propose a masking operator corresponding to dimension \textbf{0}, which helps identify and filter out fields detrimental to the target DNN model.} 
Despite the fact that neural networks can automatically adjust weights related to useless fields, the introduction of dimension {0} makes it easier to learn and observable for us to understand the importance of the field. 
{We can easily obtain the ideal $\widetilde{\mathcal{F}}$ for the retraining stage by removing the fields in $\mathcal{F}$ that are dominated by dimension {0}.}

\subsubsection{Embedding Division}
\begin{figure}[thbp!]
  \centering
  \includegraphics[width=0.5\textwidth]{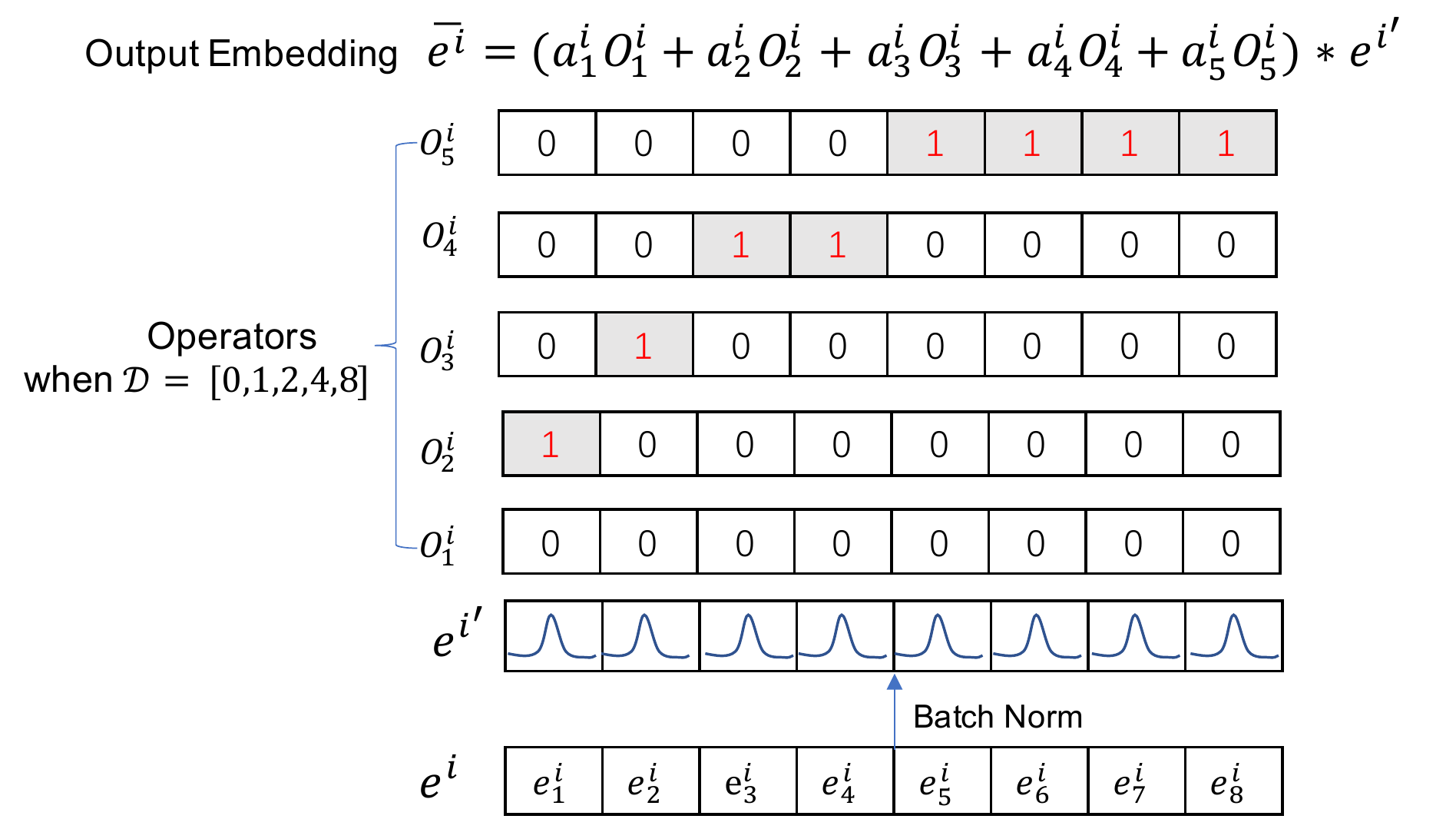}
  \caption{Embedding transformation and division in i-Razor.}
  \label{fig:embedding}
\end{figure}

In this paper, we assign the same search space to all fields for simplicity, and it is straightforward to introduce varying candidate sets. Instead of directly concatenating the $K$ candidate embedding dimensions together, a \textit{sharing embedding} architecture is introduced in order to reduce storage space and increase training efficiency. %
As shown in Figure~\ref{fig:embedding}, 
{we allocate a $d_K$-dimensional embedding to each field $f_i$, denoted by $e^i$, which is divided into $K$ independent regions. Specifically, the first region covers the front $d_1$ embedding blocks (corresponding to no dimensional blocks in the case of $d_1=0$) and the $j$-th ($1<j\leq K$) region ranges from the $(d_{j-1}+1)$-th digit to the $d_j$-th digit of $e^i$. 
To automatically learn the relative importance of each embedding region to the performance of the target DNN model,
we assign a corresponding operator $O_{j}^i \in \mathbb{R}^{d_K}$  to the $j$-th region. The elements in these operators are either 1 or 0, where {1} in a operator reflects control over the embedding block at the corresponding position, while {0} indicates the embedding block at that position is masked so that the operator is independent of that block.}  
By determining whether to keep each region in the retraining stage individually, the dimension search problem can be reformulated as an operator selection problem. 

In contrast to existing dimension search methods~\cite{zhao2020memory,zhao2020autoemb}, where different operators may share the front dimensions, the proposed decoupled structure is tailored for learning the relative importance of each independent embedding {region}, 
allowing for greater flexibility in selecting the appropriate combination of embedding regions based on the importance weights. Notice that by setting $\mathcal{D}$ subtly, such as $\left\{d_j |\ d_j = \frac{j\times(j-1)}{2},\ j\in \left[1,K\right]\right\}$ or $\left\{d_j |\ d_j = 2^{j-1},\ j\in \left[1,K\right]\right\}$, the actual dimension search space we can get is able to cover all the integers in the interval $\left[d_1, d_K\right]$ when we aim to select some of the embedding regions to determine the final dimension size after pruning unconsidered regions. For example, given $\mathcal{D}=\{0, 1, 3, 6\}$, we can get the dimension size of 5 by selecting the third region with 2 embedding blocks and the fourth region with 3 blocks.
In contrast, previous methods can only select an embedding size that belongs to $\mathcal{D}$ because the regions controlled by different operators are overlapping, and the importance of these coupled operators is not independent of each other.

\subsubsection{Continuous Relaxation}
Inspired by DARTS~\cite{liang2019darts+, liu2018darts}, we formulate the problem of joint optimization of feature selection and dimension search as an operator selection problem by incorporating continuous architecture parameters. 
Specifically, we introduce a soft layer that assigns varying weights $\widetilde{\alpha^i}=\left[\alpha_1^i,\cdots,\alpha_{K}^i\right]$
to learn the relative contribution of different embedding regions of field $f_i$. 
Trainable variables $\mathbf{w}^i=\left[w_1^i,\cdots,w_{K}^i\right]$ are introduced to calculate the value of each $\alpha^i_j$:
\begin{equation}
  \label{alph_weight}
  \alpha_j^i = \frac{exp(w_j^i/\tau)}{\sum_{j=1}^K exp(w_j^i / \tau)}, \text{ for }  j = 1,2,\ldots, K,
\end{equation} 
where $\tau>0$ is the temperature hyperparameter.
Note that the magnitude of the values in $e^i$ can be different and easily coupled with other parameters in DNN, making it difficult for the architecture parameters $\widetilde{\alpha^i}$ to represent the relative importance of different embedding regions. Following ~\cite{liu2020autofis,pin_2019}, we adopt the batch normalization~\cite{ioffe2015batch} technique before the embedding transformation to eliminate the scale issue. 
The normalized embedding $e^{i^{\prime}} = \{ e^{i^{\prime}}_1, \cdots , e^{i^{\prime}}_{d_K}\}$ is calculated as: 
\begin{equation}
  e^{i^{\prime}}_j = \frac{e^i_j - \mu^i_j(B)}{\sqrt{{[\sigma^i_j(B)]}^2+\epsilon}}, \  \text{ for }  j = 1,2,\ldots, d_K, %
\end{equation}
where $\mu^i_j(B)$ and $\sigma^i_j(B)$ are the mean and standard deviation of $e_j^i$ over a mini-batch ${B}$, and $\epsilon$ is a small positive constant to avoid numerical overflow. 
{With this trick, we can force each value in $e^{i^{\prime}}$ }to approximately follow the Gaussian distribution $\mathbb{N}(0, 1)$, making them comparable with each other.
As illustrated in Figure~\ref{fig:embedding}, the transformed embedding $\overline{e^i}$ of field $f_i$ can be computed as:
\begin{equation}
\label{equ_emb}
   \overline{e^i} = e^{i^{\prime}} \odot \sum_{j=1}^K{\alpha_{j}^i O^i_j},
\end{equation}
where $e^{i^{\prime}}$ is the output of the batch normalization layer and $\odot$ is the element-wise product. 

By applying Equation (\ref{equ_emb}) to all fields and feeding all output embedding to the target DNN structure, we can obtain the relative importance of different embedding regions in the target model with a training set. 
Such an end-to-end importance evaluation mechanism allows the DNN model to softly select different embedding regions during model training, and an ideal subset of fields $\widetilde{\mathcal{F}} \subseteq \mathcal{F}$ and the corresponding discrete mixed dimension scheme $\mathcal{C}_\mathcal{A}$ can then be derived after training.

\subsubsection{Loss Function}
In recommendation tasks, it is common for the number of features in different fields to vary greatly. For example, the field \textit{Gender} usually has two features, while the field \textit{Browsed Goods} may have millions of features. Therefore, 
{varying even one embedding dimension on fields like \textit{Browsed Goods} can lead to a significant change in the embedding lookup table size.}
In order to balance model performance and memory consumption, inspired by ~\cite{mei2019atomnas}, we propose a novel regularization term $\mathcal{L}_p$ to regularize the number of model parameters:
\begin{align}
  \mathcal{L}_{p} &= \sum_{i=1}^N \left(\frac{|f_i|}{\sum_{j=1}^N |f_j|}\times \sum_{m=1}^K \left(c_m \alpha^i_m\right)\right),\\
  c_m &=\begin{cases}\label{cum} d_1 , m=1; \\
  d_m - d_{m-1} , \text{otherwise},
\end{cases}
\end{align}
where $|f_i|$ indicates the number of features contained in field $f_i$ and $c_m$ represents the size of the embedding region that operator $O_m^i$ controls.
We intuitively expect the weighting scheme of $\mathcal{L}_p$ to penalize those fields and regions with large cardinality, thus facilitating the compression of the model.
In this paper, we focus on the CTR prediction task, where cross-entropy loss is commonly used. We combine it with $\mathcal{L}_p$ to define the loss function of the pretraining stage as follows:
\begin{equation}
  \mathcal{L}(y, \hat{y}) = -y\log{\hat{y}} -(1-y)\log(1-\hat{y}) + \lambda \mathcal{L}_p,
\end{equation}
where $y\in\{0,1\}$ is the real label, $\hat{y}\in [0,1]$ is the predicted probability of $y=1$, and the hyperparameter $\lambda$ is a coefficient to control the degree of regularization.

\subsubsection{One-level Optimization}
In this work, we reparameterize the problem of feature selection and dimension search as a structural search problem and relax the selection of different operators to be continuous and differentiable.
We slightly abuse the symbol
$\mathcal{A}=[\widetilde{\alpha^1}, \cdots, \widetilde{\alpha^N}]$ to denote all the continuous architecture parameters and $\mathcal{W}(\mathcal{A})$ to represent the downstream DNN parameters.
Different from DARTS-style works~\cite{zhao2020memory,cheng2020differentiable, liu2018darts} that use bi-level optimization to iteratively update $\mathcal{A}$ and $\mathcal{W}(\mathcal{A})$, we follow the setting of~\cite{liu2020autofis} and use one-level optimization to simultaneously train $\mathcal{A}$ and $\mathcal{W}(\mathcal{A})$ on the training set $\mathcal{T}$. In addition to being time-consuming,~\cite{liu2020autofis} empirically demonstrated that bi-level optimization might downgrade the final performance due to the accumulation of errors from multiple approximations. On the other hand, we argue that there are many causes that can affect the performance and stability of bi-level optimization, such as the size of the validation set and the switch interval during iterative training. Hence, we adopt one-level optimization for practicality and efficiency by simultaneously updating $\mathcal{A}$ and $\mathcal{W}(\mathcal{A})$ via gradient descent as follows:
\begin{equation}
  \partial_\mathcal{A}{\mathcal{L}\left(\mathcal{T}; \mathcal{A}, \mathcal{W}\left(\mathcal{A}\right)\right)}  \quad \text{and} \quad \partial_{\mathcal{W}\left(\mathcal{A}\right)}{\mathcal{L}\left(\mathcal{T}; \mathcal{A}, \mathcal{W}\left(\mathcal{A}\right)\right)}.
\end{equation}

\subsection{Retraining Stage}
\begin{algorithm}[t]
\caption{The CPT-based Pruning Algorithm}
\label{algo:our-method}
\SetAlgoLined
\KwIn{The value of $\widetilde{\alpha^i}$ after pretraining, the threshold $cpt$ ($0\leq cpt \leq 1$), and the search space $\mathcal{D} = \{d_1, \cdots , d_K \}$. }
\KwOut{a pruned dimension size $ {d}_{f_i}$ assigned to field $f_i$.}
Sort the architecture weights $\widetilde{\alpha^i}$ in the descending order: $\widetilde{\alpha^i} =  [\alpha^i_{j_1}, \cdots, \alpha^i_{j_K}]$, satisfying $\alpha^i_{j_1} \ge \alpha^i_{j_2} \ge \cdots \ge \alpha^{i}_{j_K}$;\\
$sum = 0$  \tcp*{The cumulative weight}
${d}_{f_i}$ = 0 \tcp*{The cumulative dimension size}
\For{$t \in [1,2,\ldots,K] $}{
     ${d}_{f_i} =  {d}_{f_i} + c_{j_t}$ \tcp*{$c_{j_t}$ is the embedding size of $\alpha^i_{j_t}$ calculated from Eq.(\ref{cum})}
     $sum = sum + \alpha^i_{j_t}$ \tcp*{Accumulation of dimensions}
     \If{$sum \geq cpt$}{
          \emph{\textbf{break}}\tcp*{Exit the loop and obtain the desired dimension size}
     }
  }
return $(f_i, {d}_{f_i})$\\
\end{algorithm}

After the {pretraining stage}, we can obtain the relative importance of different operators. We then seek to derive the optimal subset $ \widetilde{\mathcal{F}} \subseteq \mathcal{F}$ and the corresponding dimension configuration $\mathcal{C}_\mathcal{A}$.
Retraining the target DNN model with $\widetilde{\mathcal{F}}$ and $\mathcal{C}_\mathcal{A}$  aims to eliminate suboptimal influence since the unpicked fields in $\mathcal{F}$ and the redundant embedding dimensions are also involved during the pretraining stage.

\subsubsection{Input Configuration Derivation}
Despite the hard selector \textit{argmax} being widely used in previous DARTS-style work~\cite{liu2018darts, zhao2020memory}, we argue that it can lead to an inconsistency between the pretraining model and the retraining model, resulting in inferior performance. For instance, assume there is a field that has three candidate dimensions with corresponding weights of  $\{0.33,0.33,0.34\}$. 
It is difficult to judge which operator is better since the gap is too small, and dropping all of the relatively smaller operators can lead to a dilemma that the total weight of the dropped operators is much higher than the selected one.
Moreover, due to randomness during DNN model training, the derived optimal operators can also be unstable. To alleviate such issues, we set a \textbf{c}umulative \textbf{p}robability \textbf{t}hreshold (CPT) to flexibly adjust the minimum amount of information kept in the retraining stage, supplemented with a pruning algorithm to derive the embedding dimension ${d}_{f_i}$ for each field $f_i$. The details are presented in Algorithm \ref{algo:our-method}.
The intuition behind this is that embedding regions with higher weights contain more effective information and should be retained in preference during pruning. 

Recall that ${d}_{f_i}=0$ means that the corresponding field $f_i$ is useless and thus can be removed in the retraining stage. By applying Algorithm \ref{algo:our-method} to each field separately, we simultaneously complete feature selection and dimension search to obtain the pruned neural architecture of the retraining model.
Compared with those selection algorithms based on the \textit{argmax} operator, the proposed CPT-based pruning algorithm can generate more fine-grained embedding dimension configurations. Meanwhile, our algorithm can cover some common embedding search methods by flexibly adjusting the threshold. For example, the \textit{argmax} operator in DART-style work \cite{liu2018darts,zhao2020memory} becomes a special case with  $cpt=0$,  while $cpt = 1$ means retaining all regions. 

\subsubsection{Model Retraining}
As shown in Figure \ref{fig:framework}, we discard useless fields and redundant dimensions to redefine the architecture of the retraining model. Specifically, we only allocate embedding dimensions to fields in $\widetilde{\mathcal{F}}$ according to $\mathcal{C}_\mathcal{A}$ and concatenate all of these dimensions to feed into the subsequent DNN model. Note that as the dimension size of the embedding layer changes, the first layer of the DNN model needs to be adjusted accordingly. In addition, batch normalization is not used during retraining since there are no comparisons between different operators in each field.
All parameters in the retraining model will be trained on the training set $\mathcal{T}$ by only minimizing the cross-entropy loss.

\subsection{Complexity Analysis and Discussion}
\label{sec-com}
Similar to FM-style approaches~\cite{guo2017deepfm,rendle2010factorization, zhang2016deep}, the space complexity of i-Razor is near $\mathcal{O}(nd_K)$, where $n$ is the total number of features contained in all $N$ fields and $d_K$ is the largest candidate dimension in the search space. 
In terms of training speed, the continuous relaxation and embedding transformation in i-Razor has a complexity of $\mathcal{O}(NKd_K)$, which slows down the training when setting up candidates of large embedding dimensions. Overall, the proposed i-Razor elegantly converts the joint optimization of feature selection and dimension search into an operator selection process, which improves the flexibility 
and interpretability of feature engineering, while goes beyond the convention of ``feature selection before dimension search''. Given a suitable search space, i-Razor is efficient and conforms to the demand of recommendation scenarios.

\section{Experiments}\label{sec:offline-exp}
Extensive experiments were conducted on two well-known public datasets to answer the following research questions:
\begin{itemize}
    \item \textbf{RQ1}: How does i-Razor perform when compared to other input search methods?
    \item \textbf{RQ2}: How do key components, i.e., the setting of $d_1=0$ and the CPT-based pruning algorithm, affect the performance?
    \item \textbf{RQ3}: How do different hyperparameters affect i-Razor?
    \item \textbf{RQ4}: How does the search space setting  affect i-Razor?
    \item \textbf{RQ5}: What is the interpretability  of i-Razor?
    \item \textbf{RQ6}: What is the compatibility of i-Razor with various deep recommendation models, and can the learned configurations from one model serve as a valuable reference for others on the same dataset?
\end{itemize}

\subsection{Datasets}

\begin{table}[t]
\caption{Statistics of Evaluation Datasets}
\label{tab:dataset}
\centering
\resizebox{0.49\textwidth}{!}{
\begin{tabular}{lcccc}
\hline
Dataset & \#instances & \#features & \#fields & positive ratio \\ \hline
Criteo & $1\times 10^{8}$ & $1\times 10^{6}$ & 39 & 0.50 \\
Criteo-FG & $1\times 10^{8}$ & $1\times 10^{8}$ & 780 & 0.50 \\
Avazu & $4\times 10^{7}$ & $6\times 10^{5}$ & 24 & 0.17 \\
Avazu-FG & $4\times 10^{7}$ & $4\times 10^{7}$ & 300 & 0.17 \\
\hline
\end{tabular}}
\end{table}

Experiments were conducted on the following two public datasets:
\begin{itemize}
    \item \textbf{Criteo~\footnote{http://labs.criteo.com/downloads/download-terabyte-click-logs/}}: Criteo is a benchmark industry dataset, which contains one month of click logs. In our experiments, we use ``data 6-12'' as the training set while selecting ``day 13'' for evaluation. 
    {Due to the huge volume of data and the severe label imbalance (only 3\% positive samples)}, negative down-sampling is applied to keep the positive ratio roughly at $50\%$. 
    \item \textbf{Avazu~\footnote{http://www.kaggle.com/c/avazu-ctr-prediction}}: Avazu dataset was provided for the CTR prediction challenge on Kaggle. $80\%$ of randomly  shuffled data is allotted for training and validation while the rest $20\%$ for testing. 
\end{itemize}

To make a fair comparison between different methods, we process the two datasets following the settings~\footnote{https://github.com/Atomu2014/Ads-RecSys-Datasets} in AutoFIS~\cite{liu2020autofis} and PIN~\cite{pin_2019}.  
 Meanwhile, to demonstrate the capability of feature selection from noisy fields, we augment the number of fields in the two datasets by brute-forcing all cross fields, yielding $39 + C_{39}^2 = 780$ fields in \emph{Criteo} and $24 + C_{24}^2 = 300$ fields in \emph{Avazu}. 
We renamed the new datasets as \emph{Criteo-FG} and \emph{Avazu-FG}. Some key statistics of the aforementioned datasets are summarized in Table~\ref{tab:dataset}.

\begin{table*}[t]
\centering
\caption{Parameter Setting}
\begin{threeparttable}
\begin{tabular}{|l|c|c|c|c|}
\hline
 & \multicolumn{1}{l|}{Avazu} & \multicolumn{1}{l|}{Avazu-FG} & \multicolumn{1}{l|}{Criteo} & \multicolumn{1}{l|}{Criteo-FG} \\ \hline
\multirow{2}{*}{Common} & \multicolumn{4}{c|}{\begin{tabular}[c]{@{}c@{}}MLP=$[700 \times 5, 1]$\\ bias\_opt = ftrl\\ bias\_lr = 0.01\\ net\_opt = Adagrad\\ net\_lr = 0.02\\ L2 = 0.001\end{tabular}} \\ \cline{2-5} 
 & \multicolumn{2}{c|}{\begin{tabular}[c]{@{}c@{}}batch\_size=128\\ LN=True\end{tabular}} & \multicolumn{2}{c|}{\begin{tabular}[c]{@{}c@{}}batch\_size=500\\ LN=False\end{tabular}} \\ \hline
FDE &  \tabincell{c}{sp = {[}5, 10, $\cdots$, 50{]}} & \tabincell{c}{ sp= {[}1, 2, $\cdots$, 10{]}} & \tabincell{c}{sp= {[}5, 10, $\cdots$, 50{]}} & \tabincell{c}{sp= {[}1, 2, $\cdots$, 10{]}} \\ \hline
FTRL & \multicolumn{4}{c|}{L1=1, L2=1} \\ \hline
  {AutoField} & \begin{tabular}[c]{@{}c@{}}TopK = {[}8, 10, $\cdots$, 22{]}\end{tabular} & - & \begin{tabular}[c]{@{}c@{}}TopK = {[}15, 18, $\cdots$, 36{]}\end{tabular} & \multicolumn{1}{l|}{-} \\ \hline
  {AdaFS} & \begin{tabular}[c]{@{}c@{}}TopK = {[}8, 10, $\cdots$, 22{]}\end{tabular} & - & \begin{tabular}[c]{@{}c@{}}TopK = {[}15, 18, $\cdots$, 36{]}\end{tabular} & \multicolumn{1}{l|}{-} \\ \hline
AutoFIS & \begin{tabular}[c]{@{}c@{}}L1=0.01\\ L2=0.01\end{tabular} & - & \begin{tabular}[c]{@{}c@{}}L1=0.01\\ L2=0.01\end{tabular} & \multicolumn{1}{l|}{-} \\ \hline
AutoDim  & \tabincell{c}{sp={[}1, 2, 4, 8, 16, 30{]}} & \tabincell{c}{sp={[}1, 2, 4, 6{]}} & \tabincell{c}{sp={[}1, 2, 4, 8, 16, 25{]}} & \tabincell{c}{sp={[}1, 3, 6, 10{]}} \\ \hline
DARTS & \tabincell{c}{sp={[}1, 2, 4, 8, 16, 30{]}} & \tabincell{c}{sp={[}1, 2, 4, 6{]}} & \tabincell{c}{sp={[}1, 2, 4, 8, 16, 25{]}} & \tabincell{c}{sp={[}1, 3, 6, 10{]}} \\ \hline
i-Razor & \begin{tabular}[c]{@{}c@{}}sp = {[}0, 1, 2, 4, 8, 16, 30{]}\\ $\tau = 0.05$\\ $\lambda = 0.0001$\\ $cpt=0.45$\end{tabular} & \begin{tabular}[c]{@{}c@{}}sp = {[}0, 1, 2, 4, 6{]}\\ $\tau=0.05$\\ $\lambda = 0.0001$\\ $cpt=0.3$\end{tabular} & \begin{tabular}[c]{@{}c@{}}sp={[}0, 1, 2, 4, 8, 16, 25{]}\\ $\tau=0.1$\\ $\lambda =0.0001$\\ $cpt=0.8$\end{tabular} & \begin{tabular}[c]{@{}c@{}}sp={[}0, 1, 3, 6, 10{]}\\ $\tau=0.2$\\ $\lambda = 0.0001$\\ $cpt=0.3$\end{tabular} \\ \hline
\end{tabular}
\begin{tablenotes}
\item \textbf{LN} represents layer normalization;
\item \textbf{sp} represents the search space during the pretraining stage; 
\item \textbf{L1} and \textbf{L2} control the weights of the L1-norm and L2-norm loss respectively;
\item   {\textbf{TopK}  represents the field search space, where each element means the retention of the topK fields;}
\item $\tau$ is the temperature;
\item $\lambda$ is the parameter that controls the weight of the regularization term $\mathcal{L}_p$;
\item $cpt$ is the threshold in the retraining stage.
\end{tablenotes}
\end{threeparttable}
\label{tab:para}
\end{table*}

\subsection{Experimental Setup}
\subsubsection{Baselines}
To evaluate the effectiveness of i-Razor, we select the following state-of-the-art methods as baselines:
\begin{enumerate}
\item \textbf{Feature selection methods.}
Fixed dimension embedding (FDE) specifies a uniform dimension size for all fields via traversal search and is paired with all feature selection methods to test the effectiveness of feature selection alone. 
FTRL~\cite{mcmahan2013ad} is a widely used feature selection algorithm that incorporates L1 and L2 regularization and has excellent sparsity and convergence properties.
  {AutoField~\cite{wang2022autofield} is an AutoML framework that can adaptively evaluate the inclusion/exclusion probability for each feature. Following the experimental setup in the original paper, we evaluated model performance by keeping varying numbers of features and identified the configuration that achieved the best performance. }
  {AdaFS~\cite{lin2022adafs} employs a control network to score input data samples and adaptively select the most informative features. Mirroring AdaFS-hard~\cite{lin2022adafs} in the original study, we performed hard feature selection to retain different proportions of features in order of importance and discarded the remainder.}
AutoFIS~\cite{liu2020autofis} is a state-of-the-art method to select important feature interactions in FM-based methods~\footnote{Since AutoFIS can only select cross fields, experiments were conducted only on Avazu-FG and Criteo-FG.}. Notice that we tried different deletion ratios ranging from 0 to 1 with an interval of 0.1,  and only presented the best results with the highest AUC metric.
The feature selection performance of i-Razor is evaluated by just removing the fields dominated by dimension 0.
\item \textbf{Dimension search methods.} AutoDim~\cite{zhao2020memory} and DARTS~\cite{liu2018darts} are two state-of-the-art methods that bear the closest resemblance to i-Razor and can automatically select appropriate dimensions for different fields. Both methods perform dimension search directly on all fields without feature selection to evaluate the validity of naive dimension optimization. Unlike AutoDim~\cite{zhao2020memory} and i-Razor, different dimension candidates in DARTS~\cite{liu2018darts} do not share dimension blocks with each other. DARTS needs to allocate total $\sum_{j=1}^K D_j$ dimensions to each field in the pretraining model and relies on the \emph{argmax} operator as Autodim for dimension selection.
\item \textbf{Hybrid methods.} In addition to our approach, we compare the training mechanism of feature selection followed by dimension search, and the corresponding baselines are designated as `AutoFIS w/ AutoDim' and `AutoFIS w/ i-Razor'.
\end{enumerate}

\subsubsection{Evaluation Metrics}
We evaluate the performance of our proposed method on the classical CTR prediction task. The evaluation metrics include Area Under the ROC Curve ({AUC}), the number of selected fields ({Fields}), the total embedding dimensions used in the retraining stage ({Dims}), and the parameter quantity of all embeddings ({Params})~\footnote{Dims and Params are roughly calculated by $\sum_{i=1}^N {d}_{f_i}$ and $\sum_{i=1}^N |f_i|\times {d}_{f_i}$, respectively.}. A higher AUC indicates a better recommendation performance.
It is noteworthy that a slightly higher AUC at \textbf{0.001-level} is regarded as significant for the CTR prediction task~\cite{cheng2016wide,zhao2020memory}. We naturally introduce Dims and Fields as straightforward metrics since all methods aim to reduce model parameters through feature selection or dimension search.

\subsubsection{Implementation}
We implement all algorithms in Tensorflow~\cite{abadi2016tensorflow} and use the Adagrad~\cite{duchi2011adaptive} optimizer for gradient descent. For a fair comparison, for each baseline, we have reproduced it with reference to the original paper. 
Table~\ref{tab:para} summarizes the key parameters used in our experiments. For AutoDim, we adopt the same annealing temperature $\tau = max(0.01, 1-0.00005 \cdot t)$ provided in the original paper~\cite{zhao2020memory}. 
The code is publicly available on \href{https://github.com/YaoYao1995/i-Razor.git}{https://github.com/YaoYao1995/i-Razor.git}.

\begin{table*}[thp!]
\caption{Model performance on Avazu and Avazu-FG. $^{\star\star}$ and $^{\star}$ represent significance level $p$-value $< 0.01$ and $p$-value $<0.05$ of comparing i-Razor with the best baseline (indicated by the underlined number). RelaChan represents the relative change of i-Razor compared to the underlined baseline. $\uparrow$: the higher the better; $\downarrow$: the lower the better.}
\centering
\begin{tabular}{lcccccccc}
\toprule[1pt]
 & \multicolumn{4}{c}{Avazu} & \multicolumn{4}{c}{Avazu-FG} \\ \cmidrule(r){2-5} \cmidrule(r){6-9}
Method & AUC $\uparrow$ & Fields $\downarrow$ & Dims $\downarrow$ & Params (M) $\downarrow$ & AUC $\uparrow$ & Fields $\downarrow$ & Dims $\downarrow$ & Params (M) $\downarrow$ \\ \hline
FDE & 0.7811 & 24 & 720 & 19.36 & 0.7814 & 300 & 300 & 43.89 \\
FTRL w/ FDE & 0.7812 & 20 & 600 & 19.27 & 0.7845 & 160 & 160 & 33.60 \\
  {AutoField w/ FDE} & 0.7809 & 20 & 600 & 19.35 & - & - & - & - \\
  {AdaFS w/ FDE} & 0.7812 & 12 & 360 & 19.34 & - & - & - & - \\
AutoFIS w/ FDE & - & - & - & - & 0.7835 & 189 & 189 & 16.51 \\
i-Razor w/ FDE & 0.7812 & 16 & 480 & 19.35 & 0.7870 & 75 & 75 & 27.68 \\ \midrule
AutoDim & {\ul{0.7820}} & {\ul{24}} & {\ul{239}} & {\ul{16.63}} & 0.7842 & 300 & 588 & 184.03 \\
DARTS & 0.7815 & 24 & 211 & 17.97 & 0.7839 & 300 & 596 & 201.12 \\ \midrule
AutoFIS w/ AutoDim & - & - & - & - & 0.7836 & 189 & 393 & 47.83 \\
AutoFIS w/ i-Razor & - & - & - & - & {\ul{0.7848}} & {\ul{87}} & {\ul{159}} & {\ul{22.27}} \\
i-Razor (Ours) & \textbf{0.7823} & \textbf{16} & \textbf{122} & \textbf{12.32} & \textbf{0.7896} & \textbf{75} & \textbf{108} & \textbf{40.85} \\ \midrule
RelaChan & +0.0003$^{\star}$ & -33\% & -49\% & -26\% & +0.0048$^{\star\star}$ & -14\% & -32\% & +83\% \\ \bottomrule[1pt]
\end{tabular}
\label{result:avazu}
\end{table*}
\begin{table*}[thp!]
\caption{Model performance on Criteo and Criteo-FG. FTRL is not listed as performance degrades when any proportion of fields are deleted. $\uparrow$: the higher the better; $\downarrow$: the lower the better.}
\centering
\begin{tabular}{lrcccrccc}
\toprule[1pt]
 & \multicolumn{4}{c}{Criteo} & \multicolumn{4}{c}{Criteo-FG} \\ \cmidrule(r){2-5} \cmidrule(r){6-9} 
Method & AUC $\uparrow$ & Fields $\downarrow$ & Dims $\downarrow$ & Params (M) $\downarrow$ & AUC $\uparrow$ & Fields $\downarrow$ & Dims $\downarrow$ & Params (M) $\downarrow$ \\ \midrule
FDE & \ul{0.7994} & \ul{39} & \ul{975} & \ul{29.47} & 0.8017 & 780 & 3900 & 566.62 \\
  {AutoField w/ FDE} & 0.7985 & 33 & 825 & 21.89 & - & - & - & - \\
  {AdaFS w/ FDE} & 0.7992 & 30 & 750 & 17.53 & - & - & - & - \\
AutoFIS w/ FDE & - & - & - & - & 0.8017 & 742 & 3710 & 483.67 \\
i-Razor w/ FDE & 0.7996 & 36 & 900 & 28.81 & 0.8018 & 433 & 2165 & 307.55 \\ \midrule
AutoDim & 0.7983 & 39 & 334 & 10.11 & \ul{0.8018} & \ul{780} & \ul{2087} & \ul{261.18} \\
DARTS & 0.7983 & 39 & 289 & 9.23 & 0.8015 & 780 & 1297 & 170.33 \\ \midrule
AutoFIS w/ AutoDim & - & - & - & - & 0.8017 & 742 & 1933 & 245.18 \\
AutoFIS w/ i-Razor & - & - & - & - & 0.8017 & 570 & 1476 & 182.24 \\
i-Razor (Ours) & \textbf{0.7996} & \textbf{36} & \textbf{507} & \textbf{18.13} & \textbf{0.8019} & \textbf{433} & \textbf{1197} & \textbf{169.03} \\ \midrule
RelaChan & +0.0002 & -7\% & -48\% & -38\% & +0.0001 & -45\% & -43\% & -35\% \\ \bottomrule[1pt]
\end{tabular}
\label{result:criteo}
\end{table*}

\subsection{Overall Performance (RQ1)}
Performance improvement can be an arduous process in real-world recommender systems, so we present the best performance configuration corresponding to each baseline in Table~\ref{result:avazu} and Table~\ref{result:criteo}.
There are several observations:
\begin{itemize}
\item {Proper feature selection leads to competitive or even superior performance compared with FDE. This indicates that filtering out less informative and predictive features is beneficial for recommendation models. Additionally, we observe that the discarded fields have a small cardinality (i.e., a small number of features per field) on Avazu, which limits the savings in Params. This emphasizes the importance of further exploring dimension search techniques.}
\item {Comparing all dimension search methods with FDE, we observe that dimension search methods generally achieve more significant parameter savings. However, on Criteo, existing dimension search methods assign dimensions to each field, which fails to eliminate the interference of less predictive fields, leading to performance degradation. This highlights the necessity of feature selection.}
\item {Concerning hybrid methods, on Avazu-FG, we observe a noticeable drop in AUC for `AutoFIS w/ AutoDim' compared to AutoDim, and even larger differences in AUC between `AutoFIS w/ i-Razor' and i-Razor. On Criteo-FG, i-Razor consistently outperforms `AutoFIS w/ i-Razor' in all evaluation metrics, demonstrating the advantage of joint optimization over separate sequential optimization.}
\item {Dexterously balancing predictive performance and model compression, i-Razor surpasses top baselines. On Avazu-FG, i-Razor jointly optimizes to retain informative fields. With fewer fields and Dims, it tends to concentrate more dimensions on predictive high-cardinality fields while allocating relatively fewer dimensions to other retained fields, increasing Params yet notably boosting AUC. On the remaining three datasets compared to strong baselines, i-Razor aggressively slashes Params while maintaining or even improving AUC. Combining observations on Avazu-FG and the remaining datasets demonstrates i-Razor’s adept negotiation of predictive performance and compression with flexibility.}
\end{itemize}

In summary, our experiments underscore the significance and advantages of feature selection and dimension search. Feature selection eliminates less informative features without compromising model performance, while dimension search effectively reduces parameter quantities.  The findings emphasize the importance of integrating these optimization processes, as evidenced by the superior performance of i-Razor compared to other methods.

\begin{table*}[thp]
\caption{Effectiveness of our model components. $\uparrow$: the higher the better; $\downarrow$: the lower the better.}
\centering
\begin{tabular}{lcccccccc}
\toprule[1pt]
 & \multicolumn{4}{c}{Avazu-FG} & \multicolumn{4}{c}{Criteo-FG} \\ \cmidrule(r){2-5} \cmidrule(r){6-9} 
Method & AUC $\uparrow$ & Fields $\downarrow$ & Dims $\downarrow$ & Params (M) $\downarrow$ & AUC $\uparrow$ & Fields $\downarrow$ & Dims $\downarrow$ & Params (M) $\downarrow$\\ \midrule
i-Razor w/ 0 (1) & \textbf{0.7890} & 80 & 200 & 67.32 & \textbf{0.8019} & 599 & 1487 & 208.71 \\
i-Razor w/o 0 (2) & 0.7827 & 300 & 457 & 82.64 & 0.8015 & 780 & 2148 & 314.23 \\
argmax w/ 0 (3) & 0.7883 & \textbf{74} & \textbf{169} & \textbf{58.14} & 0.8013 & \textbf{256} & \textbf{625} & \textbf{84.11} \\
argmax w/o 0 (4) & 0.7826 & 300 & 449 & 81.50 & 0.8012 & 780 & 957 & 136.90 \\ \bottomrule[1pt]
\end{tabular}
\label{tab:ablation}
\end{table*}

\subsection{Ablation Study (RQ2)}
In i-Razor, the introduced dimension 0 in the search space plays an important role to filter out the useless fields. Meanwhile, the proposed CPT-based pruning algorithm makes it easy and flexible to derive the desired embedding configuration. To further study the role of these two components, we propose and compare the following four variants:
\begin{enumerate}
    \item `i-Razor w/ 0': using the CPT-based pruning algorithm while introducing dimension 0 into the search space;
    \item `i-Razor w/o 0': only using the CPT-based pruning algorithm;
    \item `argmax w/ 0': using the \emph{argmax} selector while introducing dimension 0 into the search space;
    \item `argmax w/o 0': only using the \emph{argmax} selector.
\end{enumerate}

To be specific, we consider two kinds of candidate dimension sets, i.e., $\mathcal{D}_1=\{0, 1, 3, 6, 10\}$ and $\mathcal{D}_2=\{1, 3, 6, 10\}$. The best results of these variants on Criteo-FG and Avazu-FG are shown in Table~\ref{tab:ablation} and we draw the following conclusions:
\begin{itemize}
    \item By comparing variant (1) with (2) and comparing variant (3) with (4), we observe that the variants with dimension 0 (w/ 0) consistently outperform those without dimension 0 (w/o 0). The reason is that the introduced dimension 0 can increase the distinction between different fields and avoid assigning dimensions to redundant fields. 
    This observation suggests that the introduction of dimension 0 can benefit recommender systems where noisy and redundant fields exist, especially in the face of thousands of fields in large-scale industrial recommender systems.
   \item By comparing variant (1) with (3) and comparing variant (2) with (4), we observe that the CPT-based pruning algorithm can achieve a performance gain at the expense of preserving more fields and parameters. Recall that the \emph{argmax}-based pruning algorithm is a special case of the CPT-based pruning algorithm (by setting $cpt = 0$), we conclude that the CPT-based pruning algorithm is more suitable for real-world recommender systems due to its flexibility to obtain effective coordination between model performance and memory overhead.
\end{itemize}

In summary, both components have their advantages, such as dimension 0 contributes to feature selection and parameter reduction, while the CPT-based pruning algorithm further helps the model to arrive at better input configurations.

\begin{figure}[tbh!]
    \centering
    \subfigure{
        \includegraphics[width=0.44\textwidth]{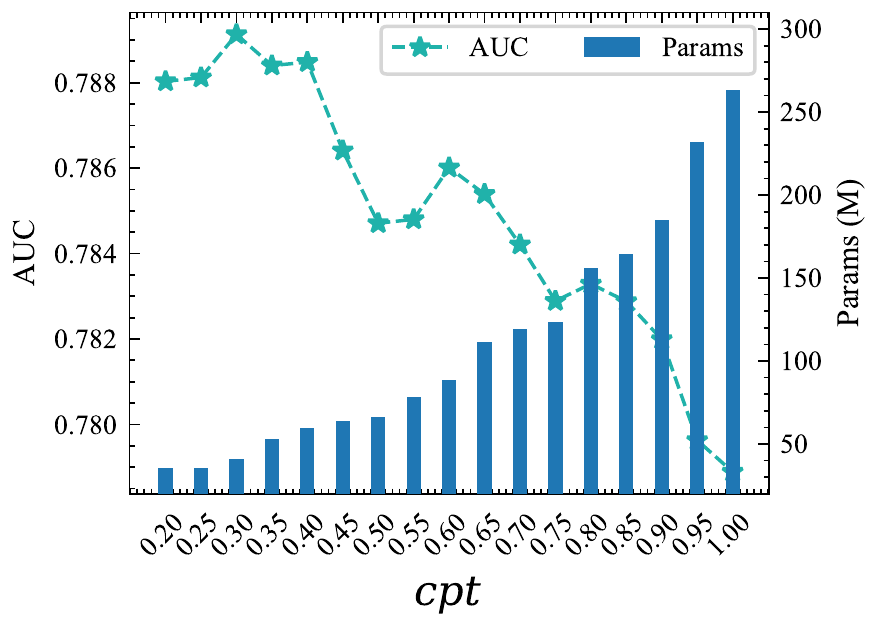}
    }
    \vspace{-3ex}
    \subfigure{
     \includegraphics[width=0.44\textwidth]{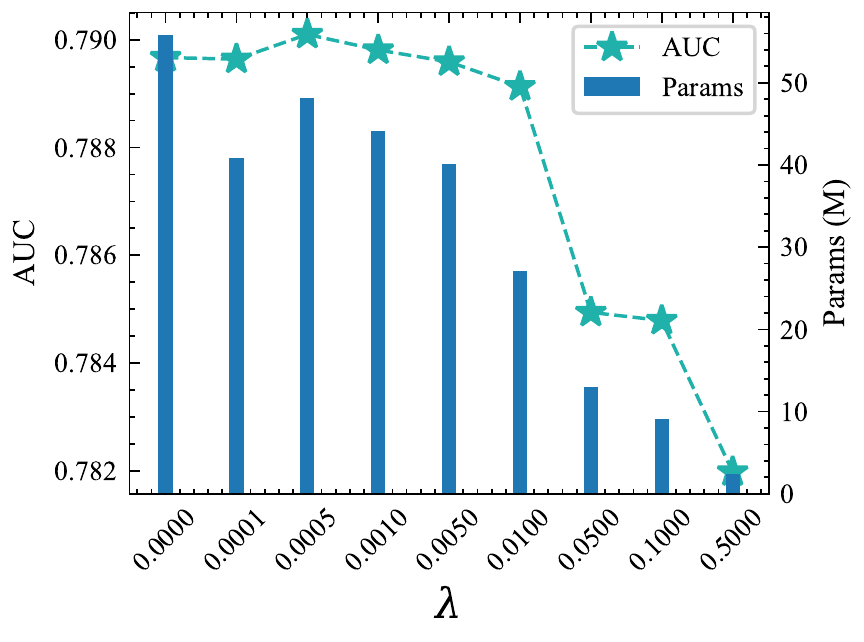}
    }
    \caption{Hyperparameter study of $cpt$ and $\lambda$ on Avazu-FG.}
    \label{fig:lamda}
    \vspace{-3ex}
\end{figure}

\subsection{Hyperparameter Study (RQ3)}
There are two main hyperparameters involved in i-Razor: the threshold $cpt$ and the weight coefficient $\lambda$ of the regularization term $\mathcal{L}_p$. 
To study the impact of $cpt$, we investigate how i-Razor performs with the change of $cpt$, while fixing other parameters ($\tau=0.05, \lambda=0.0001$). As shown in Figure~\ref{fig:lamda}, larger $cpt$ leads to larger value of Params, which is straightforward as more embedding regions and fields would be retained. %
It can be seen that when $cpt$ is relatively small ($cpt \leq 0.3$), AUC is improved with the increase of Params. However, when $cpt\geq 0.4$, the performance of the model will decrease significantly.
The potential reason is that with more fields and embedding dimensions preserved, the DNN model may require more data to eliminate the side-effect of redundant parameters. 
Meanwhile, the experimental results indicate that larger $\lambda$ usually corresponds to less Params, which reflects the effectiveness of $\mathcal{L}_p$ in controlling the number of parameters.
However, when $\mathcal{L}_p$ dominates the gradient update of the architecture parameters, model performance tends to drop sharply since some useful embedding regions or fields with large cardinality may be incorrectly filtered out.
In short, both $cpt$ and $\lambda$ are crucial hyperparameters in i-Razor, and we empirically find them to be effective and parameter insensitive when taking smaller values. In practical use, the appropriate values can be found by performing a grid search on recommended intervals: $cpt \in \left[0, 0.4\right]$ and $\lambda \in \left[0, 0.01\right]$.

\begin{figure}[tbh!]
    \centering
    \includegraphics[width=0.46\textwidth]{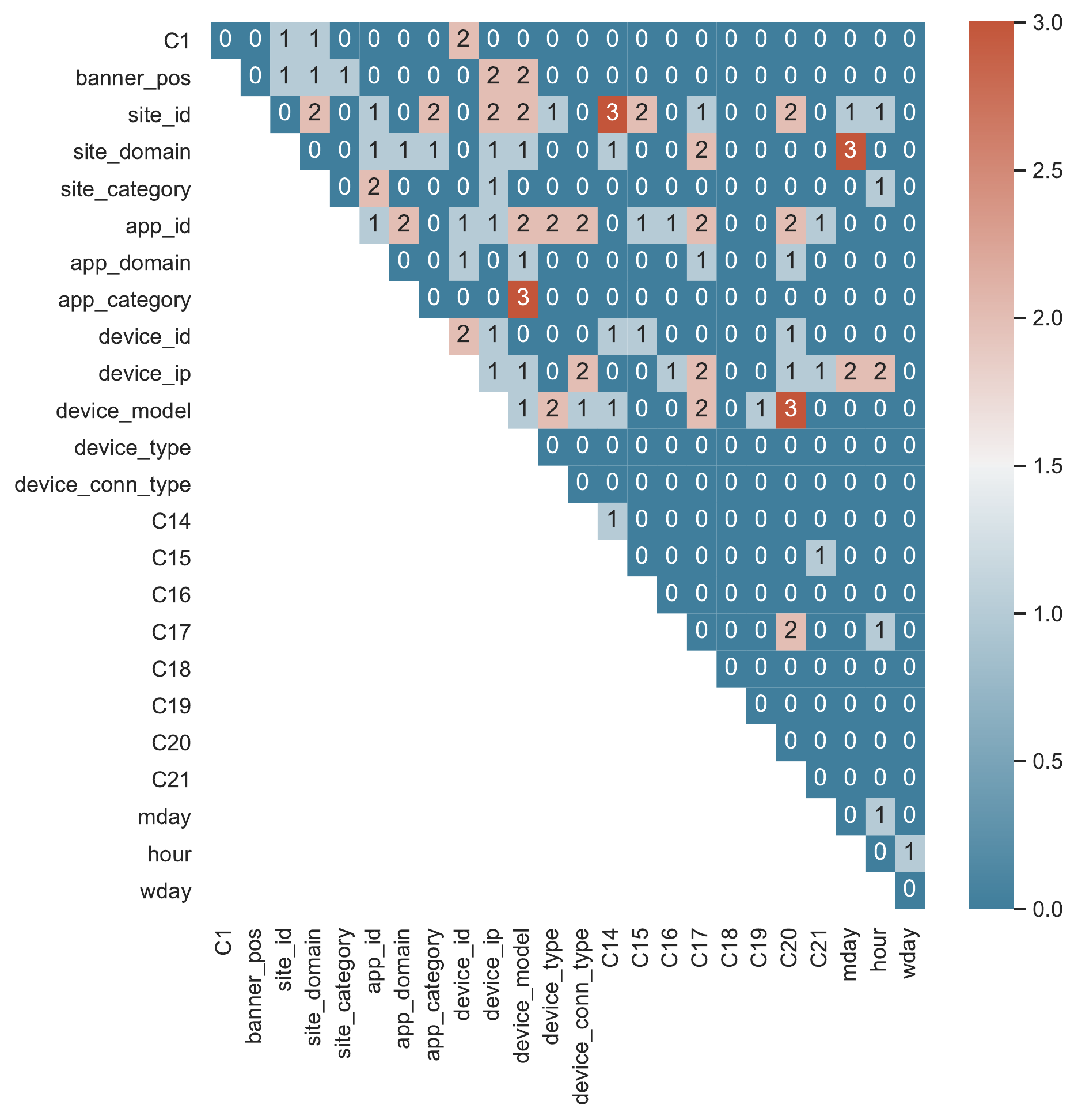}
    \caption{Embedding dimensions allocated to single fields and cross fields on Avazu-FG.}
\label{fig:dimension_visualization}
\end{figure}

\subsection{{Candidate Dimension Set Study (RQ4)}}
In this subsection, we study how different dimension partitioning schemes for the candidate dimension sets affect model performance. Specifically, we consider three heuristic settings: uniform splits $\{0, 4, 8, 12, 16, 20, 24, 28\}$, linear expansion $\{0, 1, 3, 6, 10, 15, 21, 28\}$, and exponential expansion $\{0, 1, 2, 4, 8, 16, 32\}$. 
The performance of i-Razor on Avazu under these three schemes is summarized in Table~\ref{tab:dim_config}, showing no significant differences.
This experiment provides empirical evidence that the model has a relatively low sensitivity to dimension partitioning configurations. One contributing factor is the two-stage training procedure, which separates the search for optimal configurations in pretraining from parameter tuning in retraining. The minor performance fluctuations imply the model's capability to reach similar solutions despite varying partitioning schemes.

According to the complexity analysis in Section~\ref{sec-com}, when constructing the search space by dimension partitioning, the key factors to consider are the number of candidate dimensions $K$ and the maximum dimension $d_K$.
Based on practical insights from real-world deployment, we make three recommendations for configuring the candidate dimensions: 1) Use exponential partitioning for $d_K\geq30$ to limit $K$; 2) Prefer linear expansion for $d_K\leq21$ to sufficiently cover the search space; 3) Uniform partitioning can provide coarse-grained exploration.
Meanwhile, on the Avazu and Criteo datasets, the entire two-stage training process can be completed within 3 hours and 6 hours respectively using 1 A100 40GB GPU.
This demonstrates the efficiency of i-Razor in meeting the practical training time constraints.

\begin{table}[t]
\caption{
{Impact of Candidate Dimension Set on Model Performance on Avazu.}
}
\resizebox{1\columnwidth}{!}{
\centering
\begin{tabular}{llccc}
\toprule[1pt]
       & \multicolumn{4}{c}{Metric}   \\ \cmidrule(r){2-5}
Heuristic Settings & AUC $\uparrow$ & Fields $\downarrow$ & Dims $\downarrow$ & Params (M) $\downarrow$ \\ \midrule
   Uniform Splits           & 0.7821   & 16      & 120    & 12.32      \\
   Linear Expansion         & 0.7823   & 16      & 122    & 12.32      \\
   Exponential Expansion    & 0.7822   & 16      & 125    & 12.31      \\     \bottomrule[1pt]
\end{tabular}
}
\label{tab:dim_config}
\end{table}

\subsection{Dimension Analysis (RQ5)}
We visualize the output configuration of the CPT-based pruning algorithm to get a better understanding of i-Razor. The result on Avazu-FG with the search space $\mathcal{D}=\{0,1,2,4,6\}$ is presented in Figure~\ref{fig:dimension_visualization} as an example, and similar results obtained with other settings are omitted due to limited space.

We can observe that important single fields such as ``site\_id'', ``app\_id'', and ``device\_ip'', generate many cross fields to be kept with. It makes sense since item features, context features and position features are essential features in recommender systems. For continuous fields, we can infer that ``C17'' and ``C20'' are relatively more important than others as they generate more effective cross fields. Meanwhile, we can directly pick out useful cross fields according to their derived embedding dimension size. For instance, ``site\_id * C14'', ``site\_domain * mday'', ``app\_category * device\_model'',  and ``device\_model * C20'' are expected to be beneficial cross fields, since they are assigned with the largest dimension size. 
Furthermore, we can find the configuration matrix is sparse, where only 75 out of the 300 fields are greater than 0. This means that we can directly increase the high-order interaction between different  fields to enrich the feature space, and rely on i-Razor to remove the superfluous fields. 
The visualization result also supports the conclusion that i-Razor allows  a fine-grained pruning procedure so that discriminative fields can be assigned the appropriate dimension according to their predictive capacity.
Such findings demonstrate i-Razor's capability to automatically assess the quality of fields, which facilitates the screening of new features and can guide feature engineering iterations in industrial recommender systems.

\subsection{Compatibility and Transferability Study (RQ6)}
  In this subsection, we investigate the compatibility and transferability of i-Razor.
  We examine its applicability beyond multi-layer perceptron (MLP) models~\cite{covington2016deep} and assess the effectiveness of the derived input configurations in different recommendation architectures. Our experiments focus on advanced deep recommendation models such as DeepFM~\cite{guo2017deepfm}, xDeepFM~\cite{lian2018xdeepfm},  and IPNN~\cite{qu2016product,pin_2019} using the Avazu dataset.

  To ensure dimension uniformity for feature interaction operations, we introduce different MLP layers for each field to perform linear transformations on embeddings. We compare three configurations in Table~\ref{tab:transfer}: FDE, which evenly allocates 30 dimensions to all fields; MLP, which transfers the input configuration learned from the MLP model; and model-consistent configurations specifically derived from each target model.

  Our experiments yield the following key findings:
\begin{itemize}
    \item {Directly adopting the input configuration learned from the MLP model to other deep recommendation models results in the removal of irrelevant fields, parameter compression, and improved AUC performance. These findings demonstrate the instructiveness of the MLP input configuration for enhancing the performance of other deep recommender systems on the same dataset.}
    \item {Model-consistent configurations derived specifically for each target model outperform the other two competitors. This highlights the compatibility of i-Razor with various deep recommendation models and emphasizes the importance of using model-specific configurations to achieve optimal performance.}
\end{itemize}

Overall, our experiments confirm the compatibility and transferability of i-Razor, showcasing its ability to generate input configurations that improve model performance and parameter efficiency across various deep models.

\begin{table}[t]
\caption{  {Compatibility and Transferability of i-Razor on Avazu.
$\uparrow$: the higher the better; $\downarrow$: the lower the better. $^{\star}$ indicates the statistically significant improvements (i.e.,
two-sided t-test with $p$-value $<0.05$) over the FDE config.}}
\resizebox{1\columnwidth}{!}{
\centering
\begin{tabular}{lllccc}
\toprule[1pt]
                        &        & \multicolumn{4}{c}{Metric}   \\ \cmidrule(r){3-6}
Method & Config & AUC $\uparrow$ & Fields $\downarrow$ & Dims $\downarrow$ & Params (M) $\downarrow$ \\ \midrule
\multirow{3}{*}{  {DeepFM}} & FDE    & 0.7824   & 24      & 720    & 19.36      \\
                        & MLP    & 0.7832$^{\star}$     & 16      & 122    & 12.32      \\
                        & DeepFM & 0.7835$^{\star}$   & 16      & 124    & 6.96      \\ \midrule
\multirow{3}{*}{  {xDeepFM}} & FDE  &  0.7840  & 24      & 720    & 19.36     \\
                        & MLP    & 0.7843     & 16      & 122    & 12.32  \\
                        & xDeepFM & 0.7848$^{\star}$   & 18      & 110    & 7.29      \\ \midrule
 \multirow{3}{*}{  {IPNN}}     & FDE    &  0.7852   & 24      & 720    & 19.36      \\
                        & MLP    & 0.7854   & 16      & 122    & 12.32      \\
                        & IPNN    & 0.7859$^{\star}$   & 22      & 126    & 9.46      \\     \bottomrule[1pt]
\end{tabular}
}
\label{tab:transfer}
\end{table}

\section{Conclusion}
\label{sec:conclusion}
In this paper, we address the challenge of feature selection and dimension search in modern DNN-based recommender systems.
To this end, we propose a differentiable framework that selects useful fields from a given set while assigning adaptive embedding dimensions.
The proposed method acts on the embedding layer and can be integrated into various network architectures to improve the recommendation performance and reduce the size of parameters.
The key idea is to introduce a soft selection layer that estimates the importance of different embedding regions and thus evaluates the field importance.
Moreover, we design a CPT-based pruning algorithm to flexibly generate the mixed embedding configuration.
While conceptually simple, i-Razor provides us with a concrete framework that is amenable to practicality and can be easily incorporated with various recommendation models. CTR prediction experiments on two real-world datasets verify the efficacy of i-Razor on the joint optimization of feature selection and dimension search.
Since i-Razor can serve as an efficient plug-in for recommendation models, future work may include extension to other recommendation tasks.

\ifCLASSOPTIONcompsoc
  \section*{Acknowledgments}
\else
  \section*{Acknowledgment}
\fi

This work was done during Yao Yao's internship at ByteDance Inc. 
Yao Yao and Bin Liu contributed equally to this work.
This research was also supported in part by National Natural Science Foundation of China (Grant No.71801140).

\ifCLASSOPTIONcaptionsoff
  \newpage
\fi

\bibliographystyle{IEEEtran}
\bibliography{citation}

\begin{thebibliography}{10}
\providecommand{\url}[1]{#1}
\csname url@samestyle\endcsname
\providecommand{\newblock}{\relax}
\providecommand{\bibinfo}[2]{#2}
\providecommand{\BIBentrySTDinterwordspacing}{\spaceskip=0pt\relax}
\providecommand{\BIBentryALTinterwordstretchfactor}{4}
\providecommand{\BIBentryALTinterwordspacing}{\spaceskip=\fontdimen2\font plus
\BIBentryALTinterwordstretchfactor\fontdimen3\font minus
  \fontdimen4\font\relax}
\providecommand{\BIBforeignlanguage}[2]{{%
\expandafter\ifx\csname l@#1\endcsname\relax
\typeout{** WARNING: IEEEtran.bst: No hyphenation pattern has been}%
\typeout{** loaded for the language `#1'. Using the pattern for}%
\typeout{** the default language instead.}%
\else
\language=\csname l@#1\endcsname
\fi
#2}}
\providecommand{\BIBdecl}{\relax}
\BIBdecl

\bibitem{pazzani2007content}
M.~J. Pazzani and D.~Billsus, ``Content-based recommendation systems,'' in
  \emph{The adaptive web: methods and strategies of web personalization}.\hskip
  1em plus 0.5em minus 0.4em\relax Springer, 2007, pp. 325--341.

\bibitem{chen2022clcdr}
Y.~Chen, Y.~Yao, and W.~K.~V. Chan, ``Clcdr: Contrastive learning for
  cross-domain recommendation to cold-start users,'' in \emph{International
  Conference on Neural Information Processing}.\hskip 1em plus 0.5em minus
  0.4em\relax Springer, 2022, pp. 331--342.

\bibitem{chen2023cdr}
Y.~Chen, Y.~Yao, W.~K.~V. Chan, L.~Xiao, K.~Zhang, L.~Zhang, and Y.~Ye,
  ``Cdr-adapter: Learning adapters to dig out more transferring ability for
  cross-domain recommendation models,'' \emph{arXiv preprint arXiv:2311.02398},
  2023.

\bibitem{autocross}
Y.~Luo, M.~Wang, H.~Zhou, Q.~Yao, W.-W. Tu, Y.~Chen, W.~Dai, and Q.~Yang,
  ``Autocross: Automatic feature crossing for tabular data in real-world
  applications,'' in \emph{Proceedings of the 25th ACM SIGKDD International
  Conference on Knowledge Discovery and Data Mining}, 2019, pp. 1936--1945.

\bibitem{hornik1989multilayer}
K.~Hornik, M.~Stinchcombe, H.~White \emph{et~al.}, ``Multilayer feedforward
  networks are universal approximators.'' \emph{Neural networks}, vol.~2,
  no.~5, pp. 359--366, 1989.

\bibitem{muthusankar2019high}
D.~Muthusankar, B.~Kalaavathi, and P.~Kaladevi, ``High performance feature
  selection algorithms using filter method for cloud-based recommendation
  system,'' \emph{Cluster Computing}, vol.~22, no.~1, pp. 311--322, 2019.

\bibitem{cheng2020differentiable}
W.~Cheng, Y.~Shen, and L.~Huang, ``Differentiable neural input search for
  recommender systems,'' \emph{arXiv preprint arXiv:2006.04466}, 2020.

\bibitem{chandrashekar2014survey}
G.~Chandrashekar and F.~Sahin, ``A survey on feature selection methods,''
  \emph{Computers and Electrical Engineering}, vol.~40, no.~1, pp. 16--28,
  2014.

\bibitem{wang2020bacterial}
H.~Wang, B.~Niu, and L.~Tan, ``Bacterial colony algorithm with adaptive
  attribute learning strategy for feature selection in classification of
  customers for personalized recommendation,'' \emph{Neurocomputing}, 2020.

\bibitem{zheng2023automl}
R.~Zheng, L.~Qu, B.~Cui, Y.~Shi, and H.~Yin, ``Automl for deep recommender
  systems: A survey,'' \emph{ACM Transactions on Information Systems}, 2023.

\bibitem{wang2011igf}
G.~Wang, J.~Ma, and S.~Yang, ``Igf-bagging: Information gain based feature
  selection for bagging,'' \emph{International Journal of Innovative Computing,
  Information and Control}, vol.~7, no.~11, pp. 6247--6259, 2011.

\bibitem{dash2003consistency}
M.~Dash and H.~Liu, ``Consistency-based search in feature selection,''
  \emph{Artificial intelligence}, vol. 151, no. 1-2, pp. 155--176, 2003.

\bibitem{xue2015survey}
B.~Xue, M.~Zhang, W.~N. Browne, and X.~Yao, ``A survey on evolutionary
  computation approaches to feature selection,'' \emph{IEEE Transactions on
  Evolutionary Computation}, vol.~20, no.~4, pp. 606--626, 2015.

\bibitem{liu2015pairwise}
M.~Liu and D.~Zhang, ``Pairwise constraint-guided sparse learning for feature
  selection,'' \emph{IEEE transactions on cybernetics}, vol.~46, no.~1, pp.
  298--310, 2015.

\bibitem{friedman2001greedy}
J.~H. Friedman, ``Greedy function approximation: a gradient boosting machine,''
  \emph{Annals of statistics}, pp. 1189--1232, 2001.

\bibitem{he2014practical}
X.~He, J.~Pan, O.~Jin, T.~Xu, B.~Liu, T.~Xu, Y.~Shi, A.~Atallah, R.~Herbrich,
  S.~Bowers \emph{et~al.}, ``Practical lessons from predicting clicks on ads at
  facebook,'' in \emph{Proceedings of the eighth international workshop on data
  mining for online advertising}, 2014, pp. 1--9.

\bibitem{liu2018darts}
H.~Liu, K.~Simonyan, and Y.~Yang, ``Darts: Differentiable architecture
  search,'' in \emph{International Conference on Learning Representations},
  2018.

\bibitem{joglekar2020neural}
M.~R. Joglekar, C.~Li, M.~Chen, T.~Xu, X.~Wang, J.~K. Adams, P.~Khaitan,
  J.~Liu, and Q.~V. Le, ``Neural input search for large scale recommendation
  models,'' in \emph{Proceedings of the 26th ACM SIGKDD International
  Conference on Knowledge Discovery and Data Mining}, 2020, pp. 2387--2397.

\bibitem{zhao2020memory}
X.~Zhao, H.~Liu, H.~Liu, J.~Tang, W.~Guo, J.~Shi, S.~Wang, H.~Gao, and B.~Long,
  ``Autodim: Field-aware embedding dimension searchin recommender systems,'' in
  \emph{Proceedings of the Web Conference 2021}, 2021, pp. 3015--3022.

\bibitem{zhao2020autoemb}
X.~Zhaok, H.~Liu, W.~Fan, H.~Liu, J.~Tang, C.~Wang, M.~Chen, X.~Zheng, X.~Liu,
  and X.~Yang, ``Autoemb: Automated embedding dimensionality search in
  streaming recommendations,'' in \emph{2021 IEEE International Conference on
  Data Mining (ICDM)}.\hskip 1em plus 0.5em minus 0.4em\relax IEEE, 2021, pp.
  896--905.

\bibitem{liu2021learnable}
S.~Liu, C.~Gao, Y.~Chen, D.~Jin, and Y.~Li, ``Learnable embedding sizes for
  recommender systems,'' in \emph{International Conference on Learning
  Representations}, 2020.

\bibitem{cai2018feature}
J.~Cai, J.~Luo, S.~Wang, and S.~Yang, ``Feature selection in machine learning:
  A new perspective,'' \emph{Neurocomputing}, vol. 300, pp. 70--79, 2018.

\bibitem{mafarja2020dragonfly}
M.~Mafarja, A.~A. Heidari, H.~Faris, S.~Mirjalili, and I.~Aljarah, ``Dragonfly
  algorithm: theory, literature review, and application in feature selection,''
  in \emph{Nature-Inspired Optimizers}.\hskip 1em plus 0.5em minus 0.4em\relax
  Springer, 2020, pp. 47--67.

\bibitem{lee2006information}
C.~Lee and G.~G. Lee, ``Information gain and divergence-based feature selection
  for machine learning-based text categorization,'' \emph{Information
  processing and management}, vol.~42, no.~1, pp. 155--165, 2006.

\bibitem{peng2005feature}
H.~Peng, F.~Long, and C.~Ding, ``Feature selection based on mutual information
  criteria of max-dependency, max-relevance, and min-redundancy,'' \emph{IEEE
  Transactions on pattern analysis and machine intelligence}, vol.~27, no.~8,
  pp. 1226--1238, 2005.

\bibitem{zhang2019making}
D.~Zhang, L.~Yao, K.~Chen, S.~Wang, X.~Chang, and Y.~Liu, ``Making sense of
  spatio-temporal preserving representations for eeg-based human intention
  recognition,'' \emph{IEEE transactions on cybernetics}, vol.~50, no.~7, pp.
  3033--3044, 2019.

\bibitem{1996A}
P.~E. Greenwood and M.~S. Nikulin, ``A guide to chi-squared testing,''
  \emph{Biometrics}, vol.~39, no.~4, 1996.

\bibitem{1965An}
S.~S. Shaphiro and M.~B. Wilk, ``An analysis of variance test for normality
  (complete samples),'' \emph{Biometrika}, vol.~52, pp. 591--611, 1965.

\bibitem{dudani1976distance}
S.~A. Dudani, ``The distance-weighted k-nearest-neighbor rule,'' \emph{IEEE
  Transactions on Systems, Man, and Cybernetics}, no.~4, pp. 325--327, 1976.

\bibitem{rish2001empirical}
I.~Rish \emph{et~al.}, ``An empirical study of the naive bayes classifier,'' in
  \emph{IJCAI 2001 workshop on empirical methods in artificial intelligence},
  vol.~3, no.~22, 2001, pp. 41--46.

\bibitem{ahdesmaki2010feature}
M.~Ahdesm{\"a}ki and K.~Strimmer, ``Feature selection in omics prediction
  problems using cat scores and false nondiscovery rate control,'' \emph{The
  Annals of Applied Statistics}, pp. 503--519, 2010.

\bibitem{luo2017adaptive}
M.~Luo, X.~Chang, L.~Nie, Y.~Yang, A.~G. Hauptmann, and Q.~Zheng, ``An adaptive
  semisupervised feature analysis for video semantic recognition,'' \emph{IEEE
  transactions on cybernetics}, vol.~48, no.~2, pp. 648--660, 2017.

\bibitem{pourpanah2019hybrid}
F.~Pourpanah, C.~P. Lim, X.~Wang, C.~J. Tan, M.~Seera, and Y.~Shi, ``A hybrid
  model of fuzzy min--max and brain storm optimization for feature selection
  and data classification,'' \emph{Neurocomputing}, vol. 333, pp. 440--451,
  2019.

\bibitem{chen2019semisupervised}
K.~Chen, L.~Yao, D.~Zhang, X.~Wang, X.~Chang, and F.~Nie, ``A semisupervised
  recurrent convolutional attention model for human activity recognition,''
  \emph{IEEE transactions on neural networks and learning systems}, vol.~31,
  no.~5, pp. 1747--1756, 2019.

\bibitem{cheng2022towards}
M.~Cheng, Z.~Liu, Q.~Liu, S.~Ge, and E.~Chen, ``Towards automatic discovering
  of deep hybrid network architecture for sequential recommendation,'' in
  \emph{Proceedings of the ACM Web Conference 2022}, 2022, pp. 1923--1932.

\bibitem{liu2020autofis}
B.~Liu, C.~Zhu, G.~Li, W.~Zhang, J.~Lai, R.~Tang, X.~He, Z.~Li, and Y.~Yu,
  ``Autofis: Automatic feature interaction selection in factorization models
  for click-through rate prediction,'' in \emph{Proceedings of the 26th ACM
  SIGKDD International Conference on Knowledge Discovery and Data Mining},
  2020, pp. 2636--2645.

\bibitem{wang2022autofield}
Y.~Wang, X.~Zhao, T.~Xu, and X.~Wu, ``Autofield: Automating feature selection
  in deep recommender systems,'' in \emph{Proceedings of the ACM Web Conference
  2022}, 2022, pp. 1977--1986.

\bibitem{lin2022adafs}
W.~Lin, X.~Zhao, Y.~Wang, T.~Xu, and X.~Wu, ``Adafs: Adaptive feature selection
  in deep recommender system,'' in \emph{Proceedings of the 28th ACM SIGKDD
  Conference on Knowledge Discovery and Data Mining}, 2022, pp. 3309--3317.

\bibitem{ginart2019mixed}
A.~Ginart, M.~Naumov, D.~Mudigere, J.~Yang, and J.~Zou, ``Mixed dimension
  embeddings with application to memory-efficient recommendation systems,'' in
  \emph{2021 IEEE International Symposium on Information Theory (ISIT)}.\hskip
  1em plus 0.5em minus 0.4em\relax IEEE, 2021, pp. 2786--2791.

\bibitem{chen2020differentiable}
T.~Chen, L.~Li, and Y.~Sun, ``Differentiable product quantization for
  end-to-end embedding compression,'' in \emph{International Conference on
  Machine Learning}.\hskip 1em plus 0.5em minus 0.4em\relax PMLR, 2020, pp.
  1617--1626.

\bibitem{kang2020learning}
W.-C. Kang, D.~Z. Cheng, T.~Chen, X.~Yi, D.~Lin, L.~Hong, and E.~H. Chi,
  ``Learning multi-granular quantized embeddings for large-vocab categorical
  features in recommender systems,'' in \emph{Companion Proceedings of the Web
  Conference 2020}, 2020, pp. 562--566.

\bibitem{mei2019atomnas}
\BIBentryALTinterwordspacing
J.~Mei, Y.~Li, X.~Lian, X.~Jin, L.~Yang, A.~Yuille, and J.~Yang, ``Atomnas:
  Fine-grained end-to-end neural architecture search,'' in \emph{International
  Conference on Learning Representations}, 2020. [Online]. Available:
  \url{https://openreview.net/forum?id=BylQSxHFwr}
\BIBentrySTDinterwordspacing

\bibitem{liang2019darts+}
H.~Liang, S.~Zhang, J.~Sun, X.~He, W.~Huang, K.~Zhuang, and Z.~Li, ``Darts+:
  Improved differentiable architecture search with early stopping,''
  \emph{arXiv preprint arXiv:1909.06035}, 2019.

\bibitem{franceschi2018bilevel}
L.~Franceschi, P.~Frasconi, S.~Salzo, R.~Grazzi, and M.~Pontil, ``Bilevel
  programming for hyperparameter optimization and meta-learning,'' in
  \emph{International Conference on Machine Learning}.\hskip 1em plus 0.5em
  minus 0.4em\relax PMLR, 2018, pp. 1568--1577.

\bibitem{pin_2019}
Y.~Qu, B.~Fang, W.~Zhang, R.~Tang, M.~Niu, H.~Guo, Y.~Yu, and X.~He,
  ``Product-based neural networks for user response prediction over multi-field
  categorical data,'' \emph{ACM Transactions on Information Systems (TOIS)},
  vol.~37, no.~1, pp. 1--35, 2018.

\bibitem{ioffe2015batch}
S.~Ioffe and C.~Szegedy, ``Batch normalization: Accelerating deep network
  training by reducing internal covariate shift,'' in \emph{International
  conference on machine learning}.\hskip 1em plus 0.5em minus 0.4em\relax PMLR,
  2015, pp. 448--456.

\bibitem{guo2017deepfm}
H.~Guo, R.~Tang, Y.~Ye, Z.~Li, and X.~He, ``Deepfm: a factorization-machine
  based neural network for ctr prediction,'' in \emph{Proceedings of the 26th
  International Joint Conference on Artificial Intelligence}, 2017, pp.
  1725--1731.

\bibitem{rendle2010factorization}
S.~Rendle, ``Factorization machines,'' in \emph{2010 IEEE International
  conference on data mining}.\hskip 1em plus 0.5em minus 0.4em\relax IEEE,
  2010, pp. 995--1000.

\bibitem{zhang2016deep}
W.~Zhang, T.~Du, and J.~Wang, ``Deep learning over multi-field categorical
  data,'' in \emph{European conference on information retrieval}.\hskip 1em
  plus 0.5em minus 0.4em\relax Springer, 2016, pp. 45--57.

\bibitem{mcmahan2013ad}
H.~B. McMahan, G.~Holt, D.~Sculley, M.~Young, D.~Ebner, J.~Grady, L.~Nie,
  T.~Phillips, E.~Davydov, D.~Golovin \emph{et~al.}, ``Ad click prediction: a
  view from the trenches,'' in \emph{Proceedings of the 19th ACM SIGKDD
  international conference on Knowledge discovery and data mining}, 2013, pp.
  1222--1230.

\bibitem{cheng2016wide}
H.-T. Cheng, L.~Koc, J.~Harmsen, T.~Shaked, T.~Chandra, H.~Aradhye,
  G.~Anderson, G.~Corrado, W.~Chai, M.~Ispir \emph{et~al.}, ``Wide \& deep
  learning for recommender systems,'' in \emph{Proceedings of the 1st workshop
  on deep learning for recommender systems}, 2016, pp. 7--10.

\bibitem{abadi2016tensorflow}
M.~Abadi, P.~Barham, J.~Chen, Z.~Chen, A.~Davis, J.~Dean, M.~Devin,
  S.~Ghemawat, G.~Irving, M.~Isard \emph{et~al.}, ``Tensorflow: A system for
  large-scale machine learning,'' in \emph{12th USENIX symposium on operating
  systems design and implementation (OSDI 16)}, 2016, pp. 265--283.

\bibitem{duchi2011adaptive}
J.~Duchi, E.~Hazan, and Y.~Singer, ``Adaptive subgradient methods for online
  learning and stochastic optimization.'' \emph{Journal of machine learning
  research}, vol.~12, no.~7, 2011.

\bibitem{covington2016deep}
P.~Covington, J.~Adams, and E.~Sargin, ``Deep neural networks for youtube
  recommendations,'' in \emph{Proceedings of the 10th ACM conference on
  recommender systems}, 2016, pp. 191--198.

\bibitem{lian2018xdeepfm}
J.~Lian, X.~Zhou, F.~Zhang, Z.~Chen, X.~Xie, and G.~Sun, ``xdeepfm: Combining
  explicit and implicit feature interactions for recommender systems,'' in
  \emph{Proceedings of the 24th ACM SIGKDD international conference on
  knowledge discovery and data mining}, 2018, pp. 1754--1763.

\bibitem{qu2016product}
Y.~Qu, H.~Cai, K.~Ren, W.~Zhang, Y.~Yu, Y.~Wen, and J.~Wang, ``Product-based
  neural networks for user response prediction,'' in \emph{2016 IEEE 16th
  international conference on data mining (ICDM)}.\hskip 1em plus 0.5em minus
  0.4em\relax IEEE, 2016, pp. 1149--1154.

\end{thebibliography}
\end{document}